\documentclass[sigconf]{acmart}

\usepackage{graphicx}
\usepackage{paralist}
\usepackage{multirow}
\usepackage{upgreek}
\usepackage{subcaption}
\usepackage{enumitem}
\setlist[itemize]{leftmargin=*,topsep=0em}

\usepackage{makecell}

\usepackage{amsmath}

\AtBeginDocument{%
  \providecommand\BibTeX{{%
    \normalfont B\kern-0.5em{\scshape i\kern-0.25em b}\kern-0.8em\TeX}}}

\copyrightyear{2023}
\acmYear{2023}
\setcopyright{acmlicensed}\acmConference[WWW '23]{Proceedings of the ACM
Web Conference 2023}{April 30-May 4, 2023}{Austin, TX, USA}
\acmBooktitle{Proceedings of the ACM Web Conference 2023 (WWW '23), April
30-May 4, 2023, Austin, TX, USA}
\acmPrice{15.00}
\acmDOI{10.1145/3543507.3583286}
\acmISBN{978-1-4503-9416-1/23/04}

\begin{document}
\title{Contrastive Collaborative Filtering for Cold-Start Item Recommendation}


\author{Zhihui Zhou}
\authornote{Zhihui Zhou and Lilin Zhang share the first authorship.}
\affiliation{%
  \institution{School of Computer Science}
  \country{Sichuan University, China} 
}
\email{zhouzhihui_cs@stu.scu.edu.cn}

\author{Lilin Zhang}
\authornotemark[1]
\affiliation{%
  \institution{School of Computer Science}
  \country{Sichuan University, China} 
}
\email{zhanglilin@stu.scu.edu.cn}

\author{Ning Yang}
\authornote{Ning Yang is the corresponding author.}
\affiliation{%
\institution{School of Computer Science}
  \country{Sichuan University, China}
}
\email{yangning@scu.edu.cn}


\begin{abstract}
The cold-start problem is a long-standing challenge in recommender systems. As a promising solution, content-based generative models usually project a cold-start item's content onto a warm-start item embedding to capture collaborative signals from item content so that collaborative filtering can be applied. However, since the training of the cold-start recommendation models is conducted on warm datasets, the existent methods face the issue that the collaborative embeddings of items will be blurred, which significantly degenerates the performance of cold-start item recommendation. To address this issue, we propose a novel model called Contrastive Collaborative Filtering for Cold-start item Recommendation (CCFCRec), which capitalizes on the co-occurrence collaborative signals in warm training data to alleviate the issue of blurry collaborative embeddings for cold-start item recommendation. In particular, we devise a contrastive collaborative filtering (CF) framework, consisting of a content CF module and a co-occurrence CF module to generate the content-based collaborative embedding and the co-occurrence collaborative embedding for a training item, respectively. During the joint training of the two CF modules, we apply a contrastive learning between the two collaborative embeddings, by which the knowledge about the co-occurrence signals can be indirectly transferred to the content CF module, so that the blurry collaborative embeddings can be rectified implicitly by the memorized co-occurrence collaborative signals during the applying phase. Together with the sound theoretical analysis, the extensive experiments conducted on real datasets demonstrate the superiority of the proposed model. The codes and datasets are available on https://github.com/zzhin/CCFCRec.

\end{abstract}



\maketitle

\section{Introduction}
In the era of information explosion, recommender systems have been playing a critical role for mitigating information overload in various online applications. As the most successful technique for personalized recommender systems, collaborative filtering (CF) aims at predicting items that are of interest to a specific user, based on the learned user and item embeddings capturing collaborative signals from observed user-item interactions \cite{he2017neural}. However, the interactions of new users or new items are often too sparse for reliable user/item representation learning, which leads to the so-called cold-start problem that has stubbornly plagued recommender systems for a long time \cite{schein2002methods}.

In this paper, we focus on cold-start item recommendation, i.e., recommending new items to existing users, which has aroused increasing efforts from the research community. One promising research line falls into the category of content-based methods, which usually train a generative model to project a cold item's content, e.g. attributes, text, and image, etc., onto a warm item embedding space \cite{saveski2014item,mo2015image,volkovs2017dropoutnet,Barkan2019,Li2019,Pan2019,sun2020lara,Zhu2021,chen2022generative}. For example, DropoutNet \cite{volkovs2017dropoutnet} implicitly transforms a cold-start item's content to a warm embedding with randomly dropping the warm item embeddings that are learned during the training process. MWUF \cite{Zhu2021} generates warm embeddings for cold-start items based on their features and ID embeddings with two meta networks. Recently, inspired by the impressive success of Generative Adversarial Network (GAN) \cite{goodfellow2014}, GAN-based methods have been proposed for cold-start item recommendation. For example, LARA \cite{sun2020lara} exploits an adversarial neural network with multiple generators to learn a mapping from cold-start items' attributes to user embeddings to generate virtual users for cold-start items.

\begin{figure}[t]
  \includegraphics[scale=0.35]{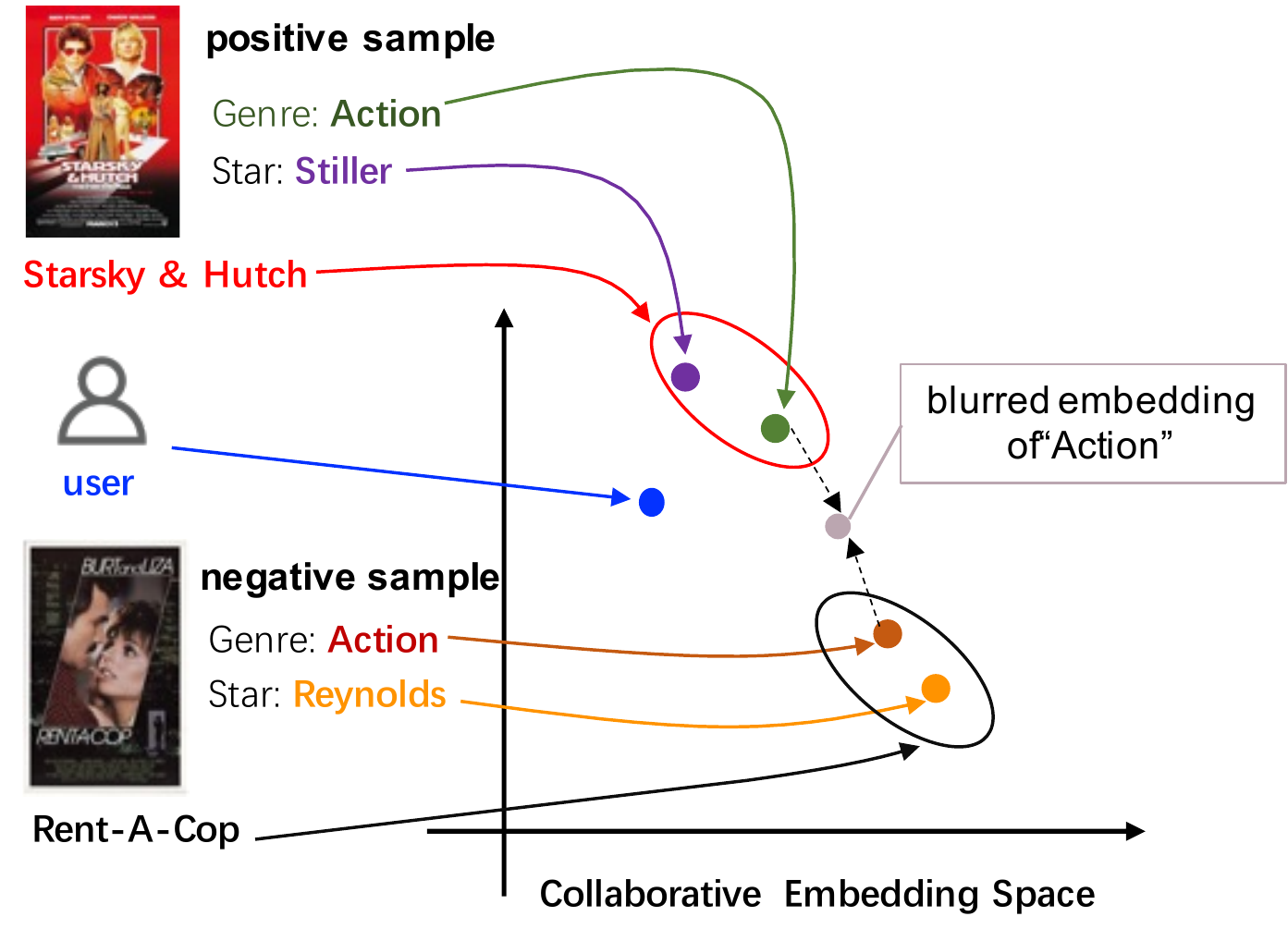}
   \caption{Illustration of blurry collaborative embedding.}
  \label{fig:motivation}
\end{figure}

Essentially, the crux of content-based methods for cold-start item recommendation is to capture collaborative signals from item content so that collaborative filtering can be applied to cold-start items. For this purpose, the existent methods often simulate the cold-start scenario during the training phase by removing the interaction records of warm items from the model input and only preserving the item content, which arguably leads to the issue of \textbf{blurry collaborative embeddings}. Figure \ref{fig:motivation} gives an illustration, where the movies Starsky \& Hutch and Rent-A-Cop are a positive sample and a negative sample of the user, respectively. During the training phase, the existent content-based methods will first embed the movie attributes, Genre and Star,  and then generate the movie's \textbf{Content-Based Collaborative Embedding (CBCE)} by aggregating its attribute embeddings. When the positive sample Starsky \& Hutch is fed into the existent models, the training algorithm will optimally push the embedding of its Genre value 'Action' to the User Collaborative Embedding (UCE) of the user, and the CBCE of Starsky \& Hutch will also be consequently close to it. In contrast, the embedding of the Genre value 'Action' of the negative sample Rent-A-Cop will be pulled away from the UCE so that the CBCE of Rent-A-Cop will be distant from it. Therefore, the final embedding of 'Action' will become blurry with the average of the 'Action' embeddings of the positive and negative samples. If in fact the user likes Action movies and did not watch Rent-A-Cop just because of the dislike of the star Reynolds, then the final blurry embedding of 'Action' will lose the user's actual preference to Action movie, which improperly pulls the positive sample's CBCE away from and pushes the negative one's CBCE close to the UCE, compared with the optimal positions where they should be located. 

On the other hand, warm datasets are affluent in co-occurrence collaborative signals of items, i.e., the patterns that some movies are frequently watched together with the same user, which are usually abandoned by the existent methods to mimic the cold-start scenarios on warm data. In sharp contrast with the existent methods, in this paper, we aim at to \textit{capitalize on the co-occurrence collaborative signals in warm data to alleviate the issue of blurry collaborative embeddings for cold-start item recommendation}. Co-occurrence collaborative signals clearly indicate that the embeddings of positive items of a user should be closer to the user's UCE than those of negative items of the same user. Therefore, if we can inject the co-occurrence collaborative signals into the CBCEs, then the blurry CBCEs can be rectified with respect to them. However, it is not trivial to achieve this goal. Unlike warm-start recommendation models which explicitly establish the mapping from historical interactions to embeddings, only item content is available to a cold-start item recommendation model, and CBCEs of cold-start items cannot explicitly and directly encode the co-occurrence collaborative signals because they are unavailable to cold-start items. 

To overcome the above challenges, in this paper, we propose a novel model called Contrastive Collaborative Filtering for Cold-start item Recommendation (CCFCRec), of which the main idea is to \textit{teach the CF module to memorize the co-occurrence collaborative signals during the training phase and how to rectify the blurry CBCEs of cold-start items according to the memorized co-occurrence collaborative signals when applying the model}. In particular, we propose a contrastive CF framework consisting of a content CF module and a co-occurrence CF module, by which CCFCRec can generate a CBCE and a \textbf{Co-Occurrence Collaborative Embedding (COCE)} for a training item, respectively. As the co-occurrence collaborative signals are unavailable for cold-start items, we present an indirect strategy to inject the co-occurrence collaborative signals into the content CF module during the training process where the co-occurrence collaborative signals are available for training items, instead of directly encoding them in CBCEs. For this purpose, CCFCRec will maximize the mutual information between CBCE and COCE of an item with an elaborate contrastive learning, which enforces the co-occurrence collaborative signals captured by COCE to be memorized by the content CF module in training phase and available in applying phase. At the same time, the two CF modules are jointly trained within a multi-task learning framework. Such multi-task optimization enforces the consistency on the interaction probability predictions that are made respectively by the content CF module and the co-occurrence CF module based on CBCE and COCE of an item, which brings the advantage of the positive transfer of the co-occurrence collaborative signals to the content CF module. At last, we also provide a sound theoretical analysis rooted in information theory to justify our contrastive collaborative filtering from the perspective of reinforcing the blurry CBCEs. The contributions of this paper are summarized as follows:

\begin{itemize}

\item We propose a novel model called Contrastive Collaborative Filtering for Cold-start item Recommendation (CCFCRec), which can rectify the blurry collaborative embeddings of cold-start items with the co-occurrence collaborative signals that are memorized during the training phase.

\item We devise a contrastive collaborative filtering framework, by which the knowledge about co-occurrence collaborative signals can be transferred to the content CF module for cold-start items.

\item We conduct extensive experiments on real datasets to verify the superiority of CCFCRec.

\end{itemize}


\section{PROBLEM FORMULATION}
Let $\mathcal{U}$ and $\mathcal{V}$ be a set of users and a set of items, respectively. Let $\mathcal{O} = \{o_{u,v} \}$ where $o_{u,v}$ is the observed interaction between a user $u \in \mathcal{U}$ and an item $v \in \mathcal{V}$. Let $\mathcal{V}_u \subseteq \mathcal{V} $ be the set of items that user $u \in \mathcal{U}$ interacted with, and $\mathcal{U}_v \subseteq \mathcal{U} $ be the set of users who interacted with item $v \in \mathcal{V}$. Each item $v$ is associated with a attribute set $\mathcal{X}_v$ consisting of $m$ attributes $ \{ x^{(v)}_1, \cdots, x^{(v)}_{m} \}$, where an attribute is represented as a one-hot vector, e.g., movie director, a multi-hot vector, e.g., movie genre, or a real valued vector, e.g., item image. Let $\mathbf{x}^{(v)}_{i} \in \mathbb{R}^{d \times 1}$ be the embedding of the $i$th attribute $x^{(v)}_{i}$, where $d$ is the dimensionality of embedding and $1 \le i \le m$.

Given a warm training dataset $\mathcal{D}$ consisting of $\{ \mathcal{U}, \mathcal{V}, \mathcal{O} \}$, we want to learn a cold-start item recommendation model to estimate the probability $\hat{y}_{u,v}$ that a user $u$ will interact with an item $v$: $\hat{y}_{u,v} = R(u, v, \mathcal{X}_v)$,
such that for any pair of $(u^{+} \in \mathcal{U}_v, u^{-} \notin \mathcal{U}_v)$ in the training dataset, $\hat{y}_{u^{+},v} > \hat{y}_{u^{-},v}$.

\section{METHODOLOGY}
  \begin{figure}[t]
	\centering
	\includegraphics[scale=0.30]{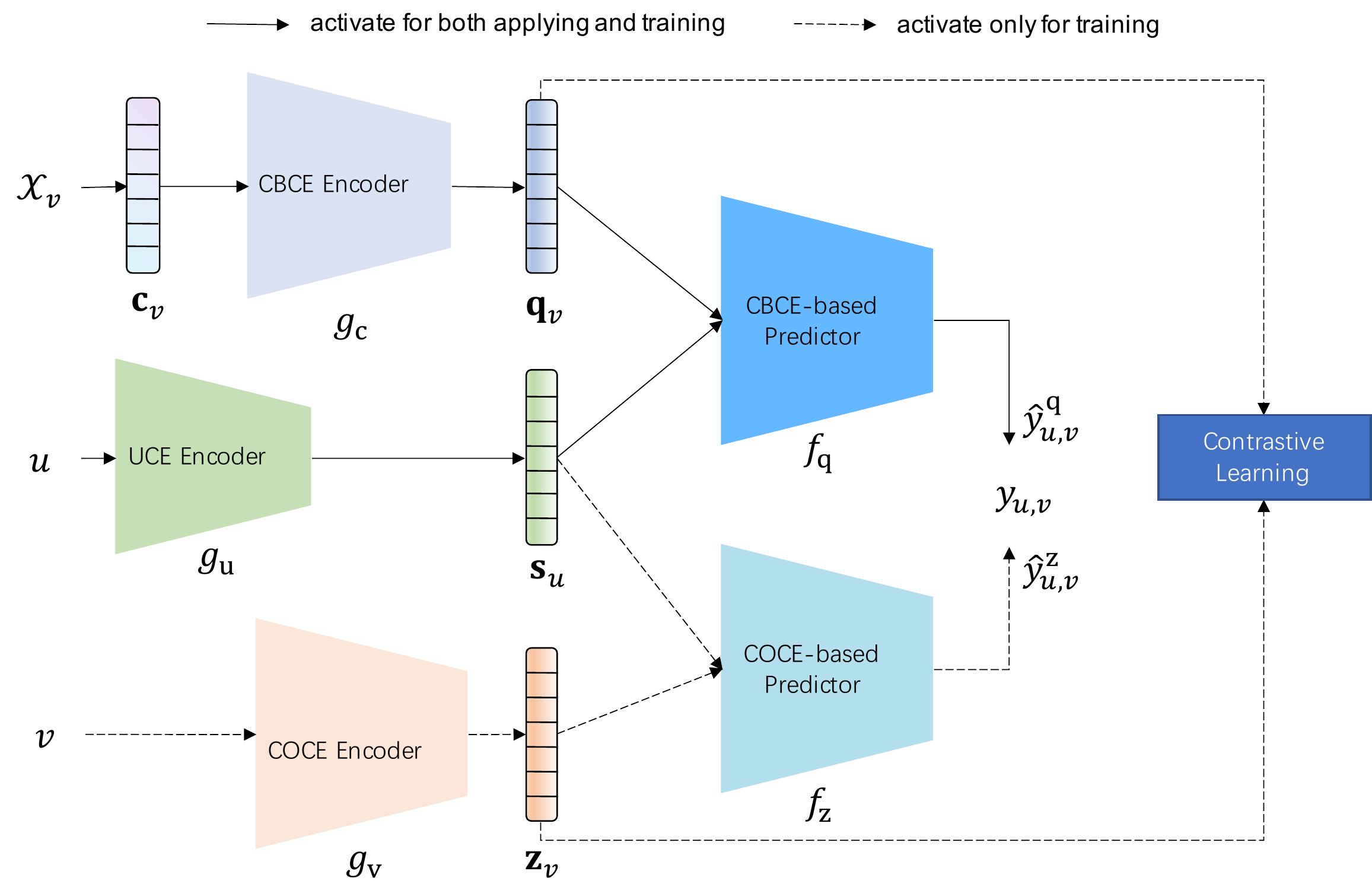}
	\caption{The overview of the proposed model CCFCRec.}
	\label{fig:model}
 \end{figure}

\subsection{Overview}

Figure \ref{fig:model} shows the architecture of CCFCRec, where the solid lines represent the paths activated when both training and applying, while the dashed lines represent the paths activated only when training. In Figure \ref{fig:model}, the content CF module includes the CBCE Encoder $g_{\text{c}}$ and the CBCE-based predictor $f_{\text{q}}$, the co-occurrence CF module includes the COCE Encoder $g_{\text{v}}$ and the COCE-based predictor $f_{\text{z}}$, and the UCE Encoder $g_{\text{u}}$ is shared between the two CF modules. 

For a training sample $(u, v,\mathcal{X}_v,$ $y_{u,v})$, CCFCRec first invokes the encoder $g_{\text{c}}$ to produce the CBCE $\mathbf{q}_{v} \in \mathbb{R}^{d \times 1}$ based on the content embedding $\mathbf{c}_v$ which is obtained by aggregating the attribute embeddings of $\mathcal{X}_v$. At the same time, CCFCRec also generates the COCE $\mathbf{z}_v \in \mathbb{R}^{d \times 1}$ of item $v$ and the UCE $\mathbf{s}_u \in \mathbb{R}^{d \times 1}$ of user $u$, via the encoders $g_{\text{v}}$ and $g_{\text{u}}$, respectively. Then CCFCRec will make the prediction of the interaction probability $\hat{y}^{\text{q}}_{u,v}$, by invoking the CBCE-based predictor $f_{\text{q}}$ with the CBCE $\mathbf{q}_{v}$ and the UCE $\mathbf{u}$ as input. Similarly, the COCE-based predictor $f_{\text{z}}$ will be invoked to make the prediction of the interaction probability $\hat{y}^{\text{z}}_{u,v}$ based on the COCE $\mathbf{z}_{v}$ and the UCE $\mathbf{s}_u$. 


To transfer the knowledge about the co-occurrence collaborative signals into the content CF module, during the training process CCFCRec will conduct a contrastive learning between $\mathbf{q}_{v}$ and $\mathbf{z}_{v}$ with optimizing objective to maximize their mutual information, which essentially captures the triple correlation $\mathbf{c}_{v} \leftrightarrow \mathbf{q}_{v} \leftrightarrow \mathbf{z}_{v}$. Although $\mathbf{z}_{v}$ is unavailable for a cold-start item when applying the model, the content CF module will memorize the triple correlation captured by the contrastive learning during the training phase and rectify a blurry $\mathbf{q}_{v}$ according to the memorized correlation during the applying phase, as if it were rectified with respect to $\mathbf{z}_{v}$ (because of the indirect correlation $\mathbf{c}_{v} \leftrightarrow \mathbf{z}_{v}$).


At last, it is noteworthy that the two CF modules are jointly trained with the same supervision signal $y_{u,v}$ in a manner of multi-task learning. As mentioned before, the joint training ensures the positive transfer of the co-occurrence collaborative signals from the co-occurrence CF module to the content CF module.

%
%

\subsection{Collaborative Embedding}

\subsubsection{CBCE} For a training item $v$, we first generate the attribute embeddings $\{ \mathbf{x}^{(v)}_i \}$, $1 \le i \le m$, which depends on how the attributes are encoded. If an attribute $x^{(v)}_i$ is represented as a one-hot or multi-hot vector, then its attribute embedding can be obtained via a lookup over a learnable embedding matrix $\mathbf{W}_i \in \mathbb{R}^{d \times |x^{(v)}_i |}$, i.e., $\mathbf{x}^{(v)}_i = \mathbf{W}_i x^{(v)}_i$. If $x^{(v)}_i$ is an image, then $\mathbf{x}^{(v)}_i$ can be obtained via a pre-trained convolutional network like VGG \cite{simonyan2014very}. Once the attribute embeddings $\{ \mathbf{x}^{(v)}_i \}$ are ready, we generate the content embedding $\mathbf{c}_v \in \mathbb{R}^{md \times 1}$ by a concatenation of $\{ \mathbf{x}^{(v)}_i \}$. At last, the CBCE $\mathbf{q}_v$ is produced via $g_{\text{c}}$ with $\mathbf{c}_v$ as input:
\begin{equation}
\mathbf{q}_v = g_{\text{c}} (\mathbf{c}_v).
\end{equation}
To capture the nonlinear interplays between attributes, we implement $g_{\text{c}}$ as a two-layer MLP with activation function LeakyReLU. 

\subsubsection{COCE and UCE}

The COCE $\mathbf{z}_v$ and UCE $\mathbf{s}_u$ encode the preference of item $v$ received from users and the preference of user $u$ offered to items, respectively, by capturing the co-occurrence collaborative signals which are hidden in the observed interactions between users and items. Therefore, we represent an item $v$ with a multi-hot vector $\mathbf{v} \in \{0, 1\}^{|\mathcal{U}| \times 1}$, and the $u$-th component $\mathbf{v}(u) = 1$ if $u \in \mathcal{U}_v$, otherwise $\mathbf{v}(u) = 0$. Similarly, a user $u$ is represented with a multi-hot vector $\mathbf{u} \in \{0, 1\}^{|\mathcal{V}| \times 1}$, and the $v$-th component $\mathbf{u}(v) = 1$ if $v \in \mathcal{V}_u$, otherwise $\mathbf{u}(v) = 0$. For simplicity, the COCE encoder $g_{\text{v}}$ and the UCE encoder $g_{\text{u}}$ are implemented as a linear combination over the columns of an embedding matrix, i.e., 
\begin{equation}
\mathbf{z}_v = g_{\text{v}}(\mathbf{v}) = \mathbf{W}_v \mathbf{v},
\end{equation}
\begin{equation}
\mathbf{s}_u = g_{\text{u}}(\mathbf{u}) = \mathbf{W}_u \mathbf{u},
\end{equation}
where $\mathbf{W}_v \in \mathbb{R}^{d \times |\mathcal{U}|}$ and $\mathbf{W}_u \in \mathbb{R}^{d \times |\mathcal{V}|}$ are the learnable embedding matrices.

\subsection{Contrastive Collaborative Filtering}

Basically, CBCE captures the preference of users to attributes of items, while COCE captures the co-occurrence collaborative signals from the observed interactions. As mentioned before, the co-occurrence collaborative signals can help alleviate the issue of blurry CBCE. However, we cannot directly encode them into a CBCE since they are available only for warm-start items during the training phase. Therefore, we take a novel strategy to teach the content CF module to memorize the co-occurrence collaborative signals in its parameters during the training phase, so that the content CF module is able to rectify the blurry CBCEs of cold-start items with respect to the memory during the applying phase. For this purpose, we regard the CBCE $\mathbf{q}_v$ and the COCE $\mathbf{z}_v$ as the content view and behavior view of an item $v$ and conduct a contrastive learning between the two item views. In particular, during the training phase, the parameters of CBCE encoder $g_{\text{c}}$ will be adjusted according to COCE so as to maximize the mutual information between the two item views. To achieve this goal, for a training item $v$, we build its positive sample set as $\mathcal{N}^{+}_v=\{ v^{+} : \mathcal{U}_v \cap \mathcal{U}_{v^+} \ne \emptyset\}$, i.e., the items with which some user interacted together with $v$, and its negative sample set $\mathcal{N}^{-}_v= \mathcal{V} \setminus \mathcal{N}^{+}_v$. To maximize the mutual information between $\mathbf{q}_v$ and $\mathbf{z}_{v^+}$, and minimize that between $\mathbf{q}_v$ and $\mathbf{z}_{v^-}$, following the idea of InfoNCE \cite{oord2018representation}, we define the contrastive loss as follow:
\begin{equation}
\mathcal{L}_{\text{c}} = - \mathbb{E}_{v \in \mathcal{D}, v^+ \in \mathcal{N}^{+}_v} \bigg[ \ln \frac{ \exp( \frac{\langle \mathbf{q}_v, \mathbf{z}_{v^+} \rangle}  {\tau}) } { \exp( \frac{\langle \mathbf{q}_v, \mathbf{z}_{v^+} \rangle}  {\tau}) + \sum_{v^{-} \in \mathcal{N}^{-}_v} \exp( \frac{\langle \mathbf{q}_v, \mathbf{z}_{v^-} \rangle}  {\tau})} \bigg],
\label{eq:Lc}
\end{equation}
where $\langle \cdot, \cdot \rangle$ represents inner product.

\subsection{Interaction Prediction}

During the training phase, CCFCRec will make two predictions of the probability of the interaction between a user $u$ and an item $v$, invoking the CBCE-based predictor $f_{\text{q}}$ and the COCE-based predictor $f_{\text{z}}$, respectively. As the predictors rank the similarity of user and item representations, we implement the predictors as inner product for its simplicity, i.e.,
\begin{equation}
\hat{y}^{\text{q}}_{u,v} = g_{\text{q}}(\mathbf{q}_v, \mathbf{s}_u) =\langle \mathbf{q}_v, \mathbf{s}_u \rangle, 
\end{equation}
\begin{equation}
\hat{y}^{\text{z}}_{u,v} = g_{\text{q}}(\mathbf{z}_v, \mathbf{s}_u) =\langle \mathbf{z}_v , \mathbf{s}_u \rangle.
\end{equation}

For a training item $v$, we pair each positive user $u^{+} \in \mathcal{V}_u$ with one negative user $u^{-}$ which is sampled from $\mathcal{V} \setminus \mathcal{V}_u$. Then by applying the popular pair-wise ranking loss BPR \cite{rendle2009bpr}, we define the following loss functions for the joint training of the two predictors:
\begin{equation}
\mathcal{L}_{\text{q}} = - \sum_{(v,u^+,u^-) \in \mathcal{D}} \ln \sigma(\hat{y}^{\text{q}}_{u^+,v}-\hat{y}^{\text{q}}_{u^-,v}),
\end{equation}
\begin{equation}
\mathcal{L}_{\text{z}} = - \sum_{(v,u^+,u^-) \in \mathcal{D}} \ln \sigma(\hat{y}^{\text{z}}_{u^+,v}-\hat{y}^{\text{z}}_{u^-,v}),
\end{equation}
where $\sigma$ is the sigmoid function.

It is noteworthy that in training phase, the two CF modules share the UCE $\mathbf{s}_u$ and are jointly trained with the same supervision. Such multi-task learning offers a consistent optimizing objective which ensures the positive transfer of the co-occurrence collaborative signals to the content CF module.

\subsection{Overall Loss Function}
The overall loss function is defined as:
\begin{equation}
\mathcal{L} = \mathcal{L}_{\text{q}} + \mathcal{L}_{\text{z}} + \lambda \mathcal{L}_{\text{c}} + \Vert \mathbf{\Theta} \Vert,
\label{eq:loss}
\end{equation}
where $\mathbf{\Theta}$ represents the learnable parameters, and $\lambda$ is a factor used for regularizing the contribution of the contrastive loss. In this paper, we choose Adam \cite{kingma2014adam} as the optimizer.

\subsection{Theoretical Analysis}
\subsubsection{Justification of Contrastive Collaborative Filtering}
Let $C$, $Q$, $U$ denote the random variables of item content embedding, item CBCE, and user, respectively. A cold-start item recommendation model can be essentially seen as a projection from item content through CBCE to user, $P(\cdot): C \rightarrow Q \rightarrow \hat{U}$, where $Q$ can be seen as a hidden variable and $\hat{U}$ is the random variable representing a predicted user. Obviously, the projection $P(\cdot)$ can be broken into two steps. The first step is to use an encoder $g(\cdot): C \rightarrow Q$ to generate CBCE based on item content, and then use a predictor $f(\cdot): Q  \rightarrow \hat{U}$ to predict the user.

Let $U$ be the clean labels (supervision signals), and then the learning of $P(\cdot)$ can be framed as a Markov chain $U\rightarrow C \stackrel{g}\rightarrow Q \stackrel{f}\rightarrow \hat{U}$ with objective to maximize the mutual information $I(\hat{U}, U)$, in terms of the theory of information bottleneck \cite{tishby2000information,Naftali2015,alemi2017}. However, the data always come together with noise. For example, in our case, the blurry collaborative embeddings can be viewed as the consequence of noisy interactions. Let $U_e$ be the labels with noise $E$. For simplicity, we suppose $U_e = U + E$ and $U \bot E$. Then the learning becomes the Markov chain $U_e\rightarrow C \stackrel{g}\rightarrow Q \stackrel{f}\rightarrow \hat{U}$ which maximizes $I(\hat{U}, U_e)$ instead of $I(\hat{U}, U)$. However, $I(\hat{U}, U_e) = I(\hat{U}, U) + I(\hat{U}, E) \ge I(\hat{U}, U)$. In other words, $I(\hat{U}, U_e)$ is the upper bound of $I(\hat{U}, U)$, and optimizing the upper bound leads to suboptimal result.

When we introduce the contrastive learning, minimizing the contrastive loss defined by Equation (\ref{eq:Lc}) is equivalent to minimizing the mutual information between negative samples \cite{oord2018representation}, i.e., $I(\hat{U}, Z^-)$, where $Z^-$ represents the negative sample variable. Now let's take a close look at Figure \ref{fig:motivation} again. From Figure \ref{fig:motivation} we can see that the noise is incurred by the negative sample since it confuses the learning of the user's preference to the genre 'Action'. Base on this observation, it is reasonable to regard $E$ as a statistics of $Z^-$. According to data processing inequality \cite{thomas2006elements}, we have $I(\hat{U}, E) \le I(\hat{U}, Z^-)$. Considering $I(\hat{U}, U) = I(\hat{U}, U_e) - I(\hat{U}, E)$, then we have $I(\hat{U}, U) \ge I(\hat{U}, U_e) - I(\hat{U}, Z^-)$. Therefore, our CCFCRec not only maximizes $I(\hat{U}, U_e)$, but also minimizes $I(\hat{U}, Z^-)$. In other words, CCFCRec maximizes the lower bound of $I(\hat{U}, U)$, which can achieve the solution closer to the optimal one.  

\subsubsection{Connection to Supervised Contrastive Learning}
The projection $P(\cdot):C \rightarrow \hat{U}$ is actually a multi-label classification task, where item content $C$ is the input and user $U$ is the label. Essentially, the contrastive CF selects the positive and negative views for an item with respect to label information, i.e., if $\mathcal{U}_v \cap \mathcal{U}_{v'} \ne \emptyset$, then $v'$ is the positive view of $v$, otherwise the negative view. As the supervision signals are leveraged for the building of contrastive views, the contrastive CF can be regarded as a Supervised Contrastive Learning (SCL) \cite{khosla2020supervised}. In fact, the contrastive loss $\mathcal{L}_{\text{c}}$ defined in Equation (\ref{eq:Lc}) can be rewritten as follow:

\begin{equation}
\begin{aligned}
\mathcal{L}_{\text{c}} = - \mathbb{E}_{v \in \mathcal{D}, v^+ \in \mathcal{N}^{+}_v} \sum_{v'\in \mathcal{D}}\mathbb{I}(v \ne v') \mathbb{I}(\mathcal{U}_{v} \cap \mathcal{U}_{v'} \neq \emptyset) f(v, v'),
\end{aligned}
\end{equation}
where $f(v, v') = \ln \bigg[ \frac{ \exp( \frac{\langle \mathbf{q}_v, \mathbf{z}_{v'} \rangle}  {\tau}) } {\sum_{v \ne  v'} \exp( \frac{\langle \mathbf{q}_v, \mathbf{z}_{v'} \rangle}  {\tau})} \bigg]$, and $\mathbb{I}(x) = 1$ if $x$ is true, otherwise $\mathbb{I}(x) = 0$. Therefore, the label information ($\mathcal{U}_{v}$ and $\mathcal{U}_{v'}$) implicitly plays the role of supervision signals for the contrastive CF, which further helps the contrastive CF distinguish the positive sample embeddings from the negative ones.



\section{Experiments}

The experiments aim to answer the following research questions:

\begin{itemize}
\item \textbf{RQ1} How does CCFCRec perform as compared to the state-of-the-art cold-start item recommendation models?
\item \textbf{RQ2} How do the contrastive CF and the joint training contribute to the performance of CCFCRec?
\item \textbf{RQ3} How do the hyper-parameters affect the performance of CCFCRec?
\item \textbf{RQ4} How can the superiority of CCFCRec be illustrated with intuitive and visualizable case studies?
\end{itemize}

\subsection{Experimental Settings}
\subsubsection{Datasets}
We conduct the experiments on the following two public datasets: MovieLens-20M (ML-20M)\footnote{https://grouplens.org/datasets/movielens/} and Amazon Video Games (Amazon-VG)\footnote{https://jmcauley.ucsd.edu/data/amazon/}. The statistics of the datasets are given in Table \ref{tbl:dataset}. On each dataset, we randomly select 70\% of the data as training set, 15\% as validation set, and 15\% as testing set. The raw embeddings of the images in the datasets are obtained through pre-trained model VGG \cite{simonyan2014very}. 
%

\begin{table}
	\begin{tabular}{ccccc}
    \toprule
    Dataset & \#Interactions & \#Users & \#Items &Sparsity  \\
    \midrule
    ML-20M & 19,904,260 & 138,493 & 24,003 & 0.598\%\\
    Amazon-VG & 475,952 & 52,965 & 35,322  & 0.025\%\\
  \bottomrule
      \end{tabular}
      \caption{Statistics of the experimental dataset.}
	  \label{tbl:dataset}
\end{table}

\begin{table*}[!t]
	\begin{tabular}{clllllll}
    \toprule
    Dataset & Baseline & HR@5 & HR@10  & HR@20 & NDCG@5 & NDCG@10  & NDCG@20\\
    \midrule
    \multirow{7}*{ML-20M}& NFM & 0.2119 & 0.1822 & 0.1531 & 0.3683 & 0.3842 & 0.3921\\
                 & LARA & 0.2425 & 0.2165 & 0.1829 & 0.4595 & 0.4541 & 0.4580\\
                & MTPR  & 0.2701 & 0.2393  & 0.2064 & 0.4504 & 0.4588 & 0.4721\\                
                
                & CVAR   & 0.2363 & 0.2188 & 0.1877 & 0.4301 & 0.4359 & 0.4448 \\
                & MvDGAE & \underline{0.2789} & \underline{0.2453} & \underline{0.2128} & 0.4586 & 0.4660 & 0.4720 \\
                & CLCRec & 0.2677 & 0.2371 & 0.1971 & \underline{0.4695} & \underline{0.4820} & \underline{0.4873}\\
                & CCFCRec    & \textbf{0.2969*} & \textbf{0.2592*} & \textbf{0.2230*} & \textbf{0.4798*} & \textbf{0.4933*} & \textbf{0.4962*} \\
                & Improv. & 6.06\% & 5.36\% &  4.57\% & 2.15\% & 2.29\% &1.79\% \\
    \bottomrule
    \multirow{7}*{Amazon-VG}& NFM  & 0.0162 & 0.0116 & 0.0089 & 0.0462 & 0.0532 & 0.0612\\
                                     & LARA & 0.0140 & 0.0074 & 0.0044 & 0.0370 & 0.0381 & 0.0400 \\
                                     & MTPR & 0.0161 & 0.0112 & 0.0083 & 0.0457 & 0.0518 & 0.0587 \\
                                    
                                    & CVAR  & 0.0149 & 0.0109 & 0.0078 & 0.0405 & 0.0481 & 0.0547 \\
                                    & MvDGAE& 0.0161 & 0.0126 & 0.0089 & 0.0456 & 0.0539 & 0.0604 \\
                                    & CLCRec& \underline{0.0229} & \underline{0.0189} & \underline{0.0150} & \underline{0.0646} & \underline{0.0769} & \underline{0.0879} \\
                                    & CCFCRec   & \textbf{0.0326*} & \textbf{0.0260*} & \textbf{0.0201*} & \textbf{0.0916*} & \textbf{0.1074*} & \textbf{0.1202*} \\
                                    & Improv. & 29.75\% & 27.30\% &  25.37\% & 29.48\% & 28.40\% &26.87\% \\
    \bottomrule
    \end{tabular}
    \caption{Performance comparison. The best runs per metric are marked in boldface. The best runs per metric among baseline methods are underlined. * represents the statistical significance with $p$-value <0.01. Improv. is the percentage by which CCFCRec improves the performance of the best baseline methods.}
    \label{tbl:performance}
\end{table*}

\subsubsection{Baseline Methods}
We compare CCFCRec with the following cold-start item recommendation methods:

\begin{itemize}
\item \textbf{NFM} \cite{he2017neuralfm} is a model for sparse data prediction, which combines the linearity of FM in modeling second-order feature interactions and the non-linearity of neural networks in modeling higher-order feature interactions.

\item \textbf{LARA} \cite{sun2020lara} is a cold-start recommendation model based on GAN, which jointly models user profiles from different attributes of the item using an adversarial neural network with multiple generators.

\item \textbf{MTPR} \cite{du2020learn} is a generic learning framework, which optimizes the dual item representations consisting of normal and counterfactual representations so that the training-testing discrepancy can be solved for cold-start items.

\item \textbf{MvDGAE} \cite{zheng2021multi} is a cold-start recommendation model based on Heterogeneous Information Networks. MvDGAE treats the cold-start problem as a missing data problem and employs a denoise graph auto-encoder framework, which randomly drops out some user-item interaction views in the encoders but forces the decoders to reconstruct the full view information.

\item \textbf{CVAR} \cite{zhao2022improving} employs latent variables to encode a distribution over item content information and warms up ID embeddings of cold-start items with a conditional decoder.

\item \textbf{CLCRec} \cite{wei2021contrastive} is another cold-start recommendation model based on contrastive learning, which enhances the behavioral features of cold-start items with a contrastive learning between an item's behavior view and content view.
\end{itemize}

\subsubsection{Evaluation Protocols} 
We adopt the widely used metrics Hit Ratio (HR) and Normalized Discounted Cumulative Gain (NDCG) to evaluate the performance of CCFCRec and the baseline methods. In particular, for a testing item $v$, we generate a ranking user list $l_v$, and the HR@$k$ of $v$ is $\frac{ \sum_{u \in \mathcal{U}_v }\mathbb{I}\big(rank(u, l_v) \le k \big)  } {k}$, where $rank(u, l)$ returns the rank of user $u$ in list $l$, and $\mathbb{I}(x) = 1$ if $x$ is true, otherwise $\mathbb{I}(x) = 0$. Then the average HR@$k$ is defined as
\begin{equation}
\text{HR@}k = \frac{1}{|\mathcal{D}_t|} \sum_{v \in \mathcal{D}_t}\frac{ \sum_{u \in \mathcal{U}_v }\mathbb{I}\big(rank(u, l_v) \le k \big)  } {k},
\end{equation}
and the NDCG@$k$ is defined as
\begin{equation}
\text{NDCG@}k = \frac{1}{|\mathcal{D}_t|} \sum_{v \in \mathcal{D}_t} \frac{1}{| \mathcal{U}_v |} \sum_{u \in \mathcal{U}_v} \frac{\mathbb{I}\big(rank(u, l_v) \le k \big) } { \log \big(1+rank(u) \big)}.
\end{equation}

\subsubsection{Parameter Setting} 
During the training of CCFCRec, we set the batch size to 1,024 and the initial learning rates to 5e-6 and 1e-4 on ML-20M and Amazon-VG, respectively. The embedding dimensionality $d$ is set to 128 and 256 on ML-20M and Amazon-VG, respectively. To speed up the contrastive learning process, for each training item in ML-20M, we sample 10 positive samples and 40 negative samples to form 400 positive-negative sample pairs in total, while for each training item in Amazon-VG, we sample 5 positive samples and 40 negative samples to form 200 positive-negative sample pairs in total. The balance factor of contrastive loss $\lambda$ is set to 0.5 and 0.6 on ML-20M and Amazon-VG, respectively. The parameter $\tau$ in Equation (\ref{eq:Lc}) is set to 0.1. For fairness, the hyper-parameters of the baseline methods are set to their optimal configuration tuned on validation sets.

\subsection{Performances Comparison (RQ1)}
The results of the performance comparison between CCFCRec and the baseline methods are shown in Table \ref{tbl:performance}, from which we can make the following observations and analyses:
\begin{itemize}
\item On both datasets, CCFCRec consistently outperforms the baseline methods in terms of all the metrics with $p$-value less than 0.01, which verifies the superiority of CCFCRec in cold-start item recommendation due to its contrastive CF. The contrastive CF enables CCFCRec to rectify the blurry collaborative embeddings by transferring the co-occurrence collaborative signals into the content CF module when training, which enhances the collaborative embeddings of cold-start items when applying.

\item The performances of all the methods degrade on Amazon-VG compared with those on ML-20M, which shows that the extreme sparsity of Amazon-VG impacts the learning of high-quality collaborative embeddings.

\item Surprisingly, even on the sparser dataset Amazon-VG, CCFCRec achieves more than 25\% improvement, which is much greater than on ML-20M. This result demonstrates CCFCRec's better adaptability to sparse training datasets. For a training item, the contrastive CF of CCFCRec takes advantage of both the content views and behavior views of items, which is consequently conducive to alleviate the sparsity problem.

\item Basically, the contrastive learning based methods (CLCRec and our CCFCRec) perform better than the other methods on both datasets, since the contrastive learning introduces more information to help learn stronger collaborative embeddings. However, CCFCRec is obviously superior to CLCRec, especially on Amazon-VG. This is because compared with CLCRec which limits the contrastive learning to the views of a training item $v$ itself, CCFCRec extends the contrastive scope to the second-order neighbors (i.e., the items with which some user interacted together with $v$) of a training item.

\end{itemize}

\begin{figure}[!t]
	\centering
	\begin{subfigure}{0.22\textwidth}
		\includegraphics[width=\textwidth]{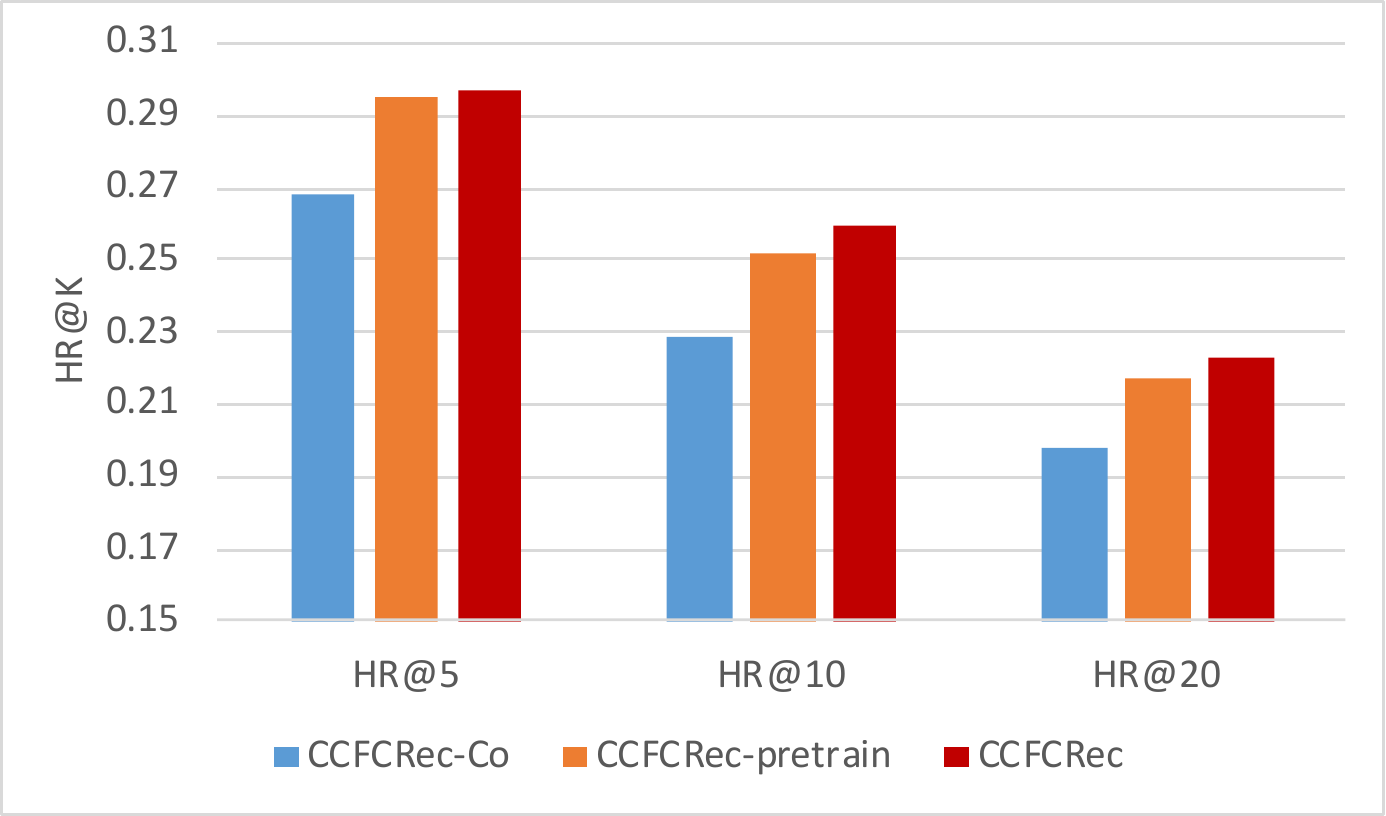}
		\caption{HR on ML-20M}
		\label{fig:HR-ML}
	\end{subfigure}
	\begin{subfigure}{0.22\textwidth}
		\includegraphics[width=\textwidth]{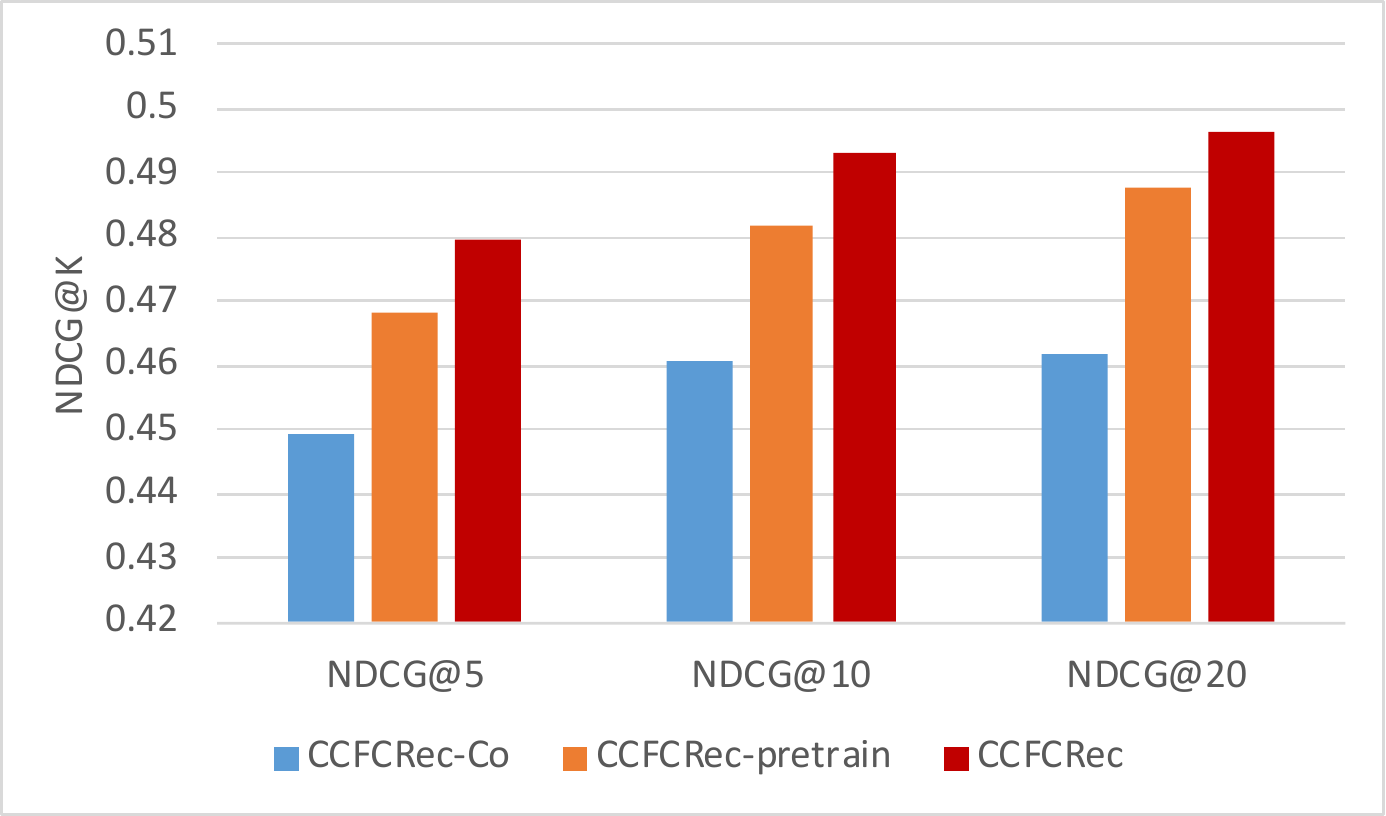}
		\caption{NDCG on ML-20M}
		\label{fig:NDCG-ML}
	\end{subfigure}
	\begin{subfigure}{0.22\textwidth}
		\includegraphics[width=\textwidth]{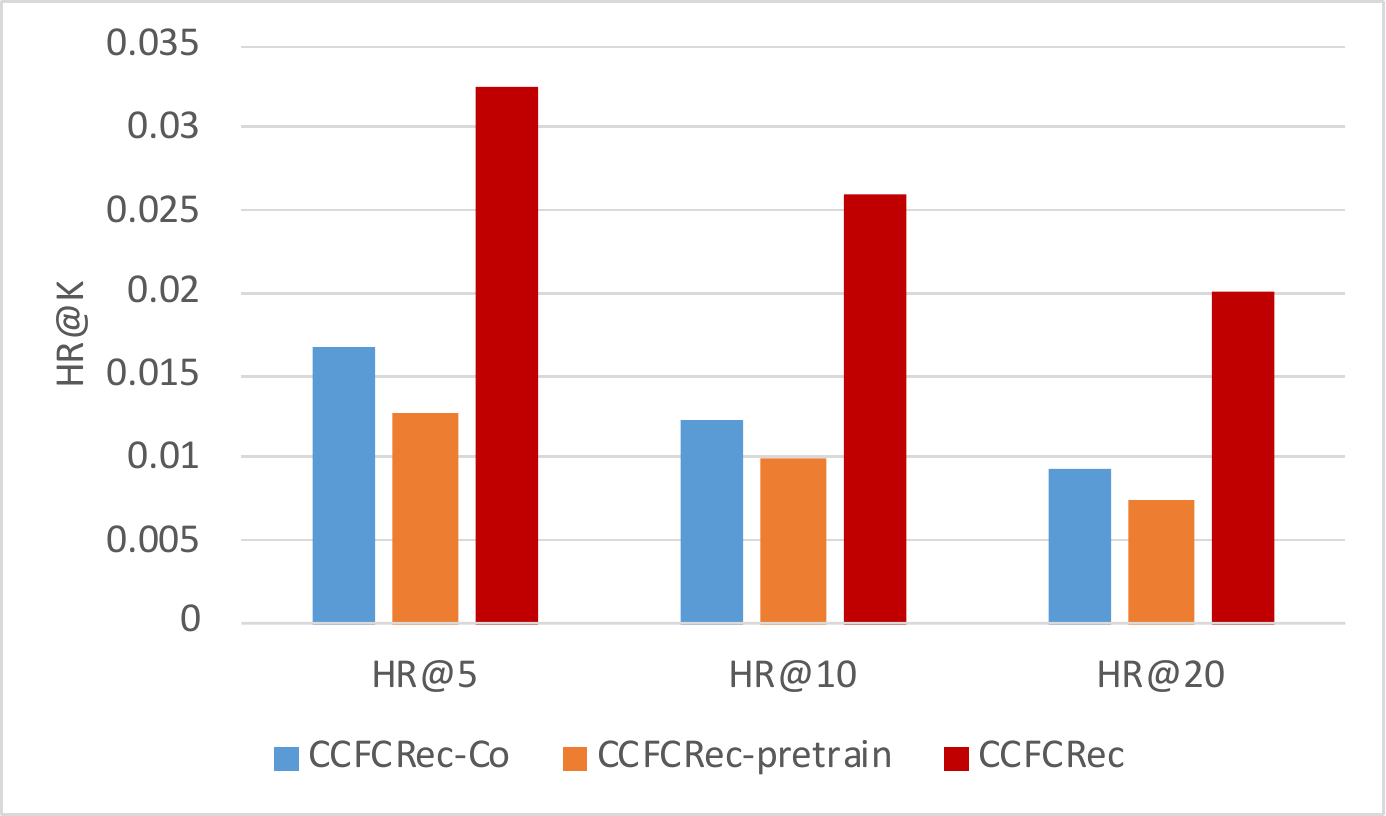}
		\caption{HR on Amazon-VG}
		\label{fig:HR-Amazon}
	\end{subfigure}
	\begin{subfigure}{0.22\textwidth}
		\includegraphics[width=\textwidth]{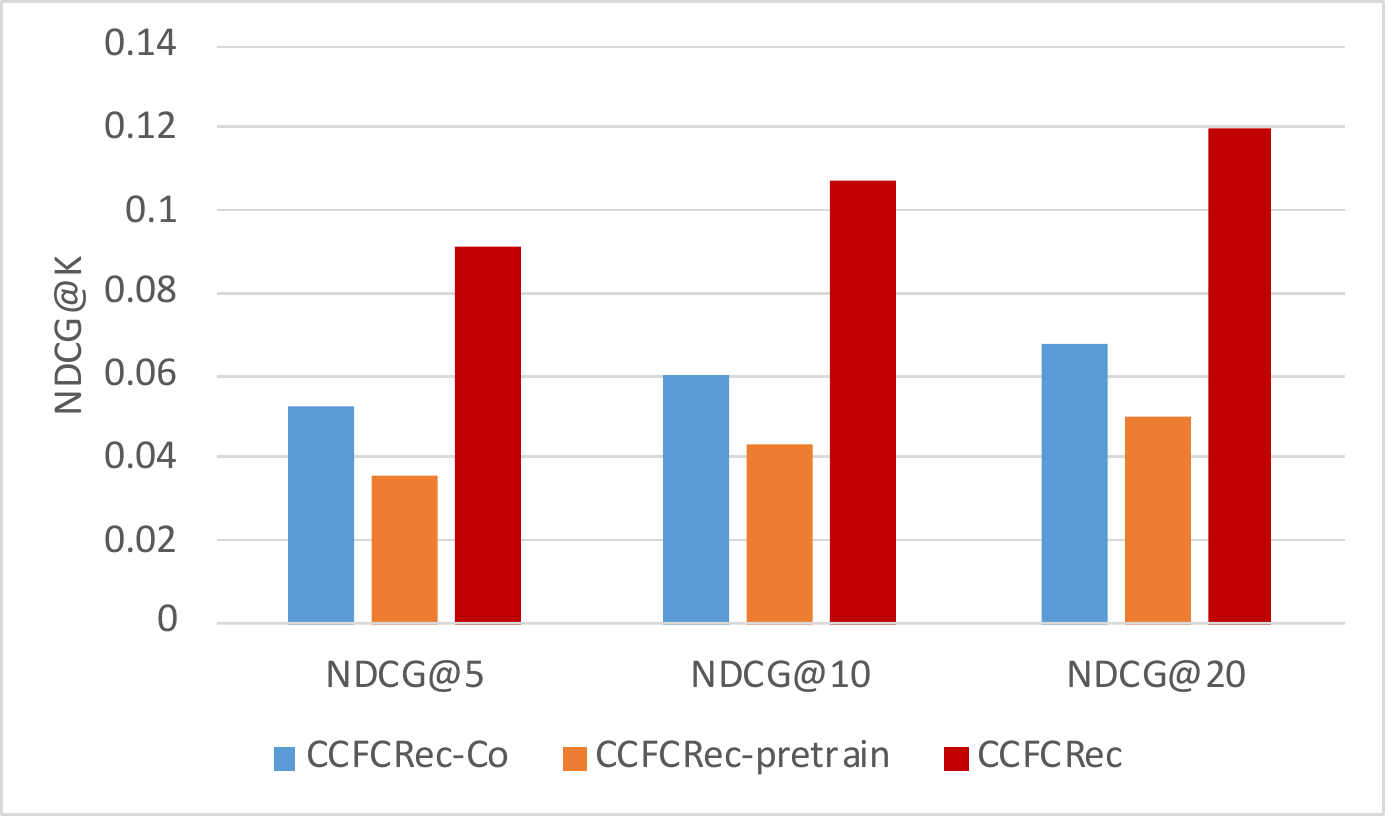}
		\caption{NDCG on Amazon-VG}
		\label{fig:NDCG-Amazon}
	\end{subfigure}
	\caption{Ablation experiments.}
	\label{fig:ablation}
\end{figure}

\begin{figure}[ht!]
	\centering
	\begin{subfigure}{0.22\textwidth}
		\includegraphics[width=\textwidth]{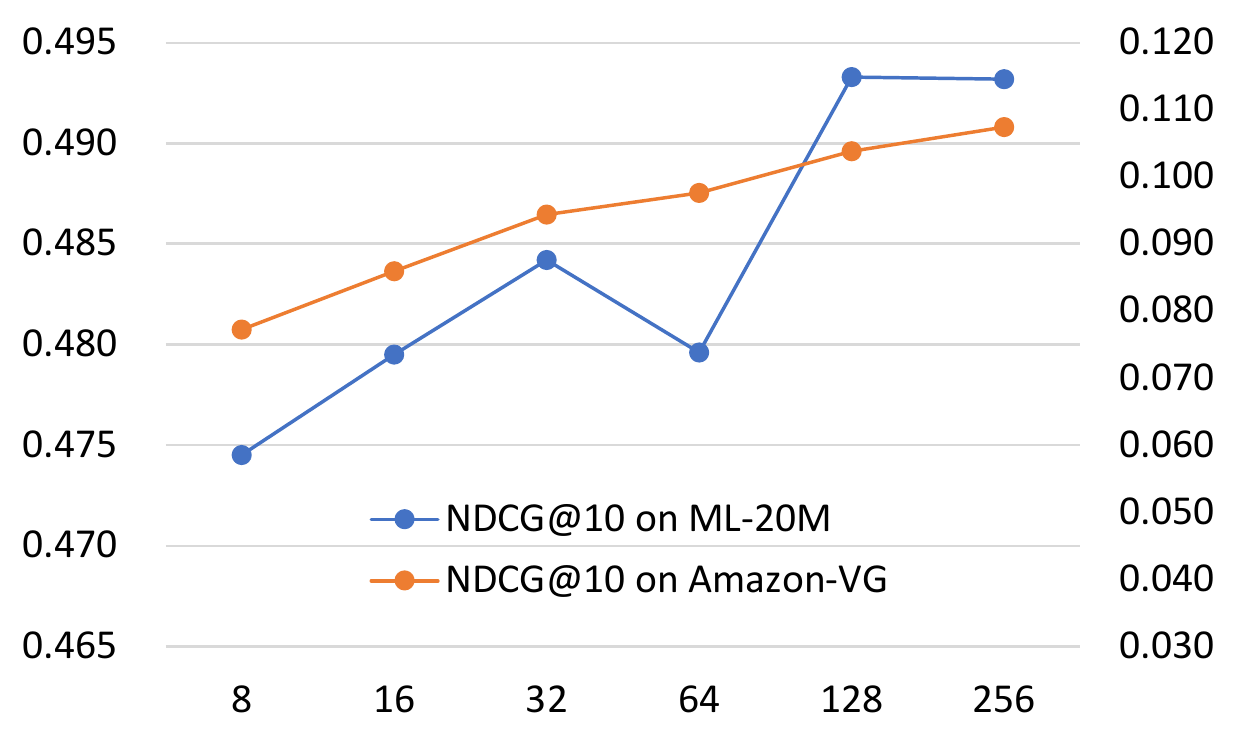}
		\caption{$d$}
		\label{fig:embeddingDimension}
	\end{subfigure}
	\begin{subfigure}{0.22\textwidth}
		\includegraphics[width=\textwidth]{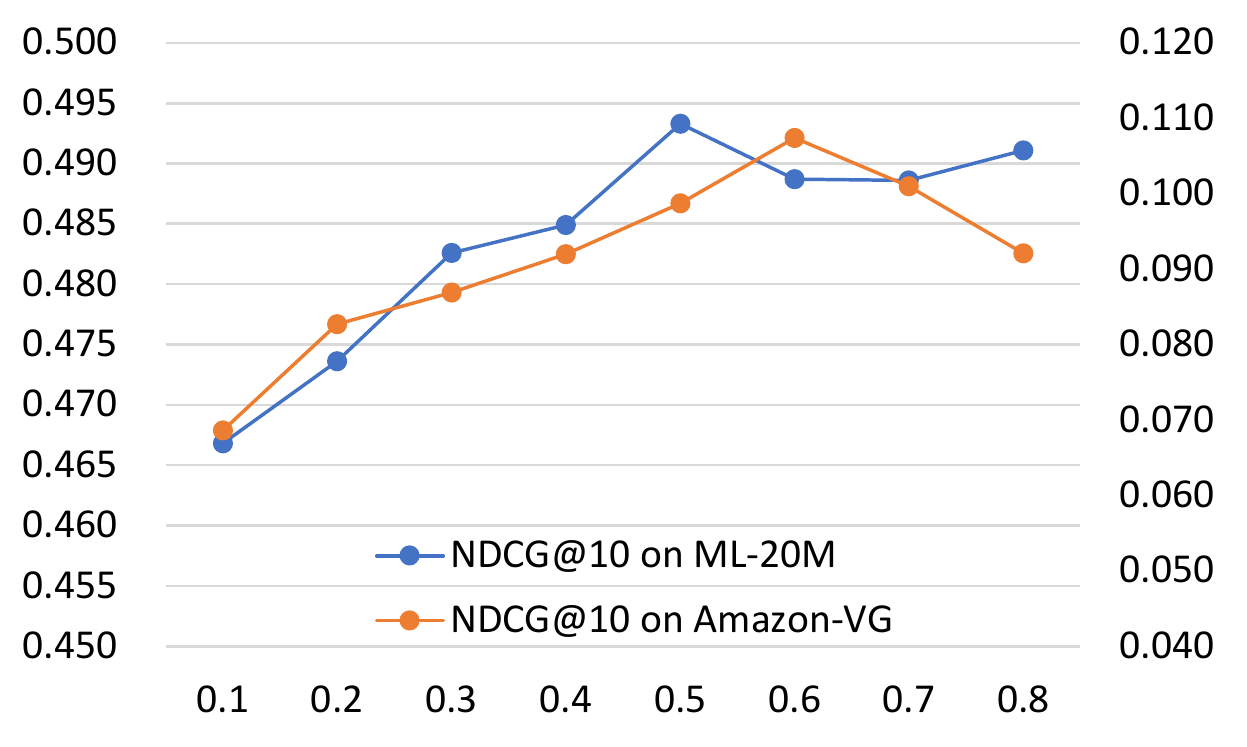}
		\caption{$\lambda$}
		\label{fig:weightOfContrastive}
	\end{subfigure}
	\begin{subfigure}{0.22\textwidth}
		\includegraphics[width=\textwidth]{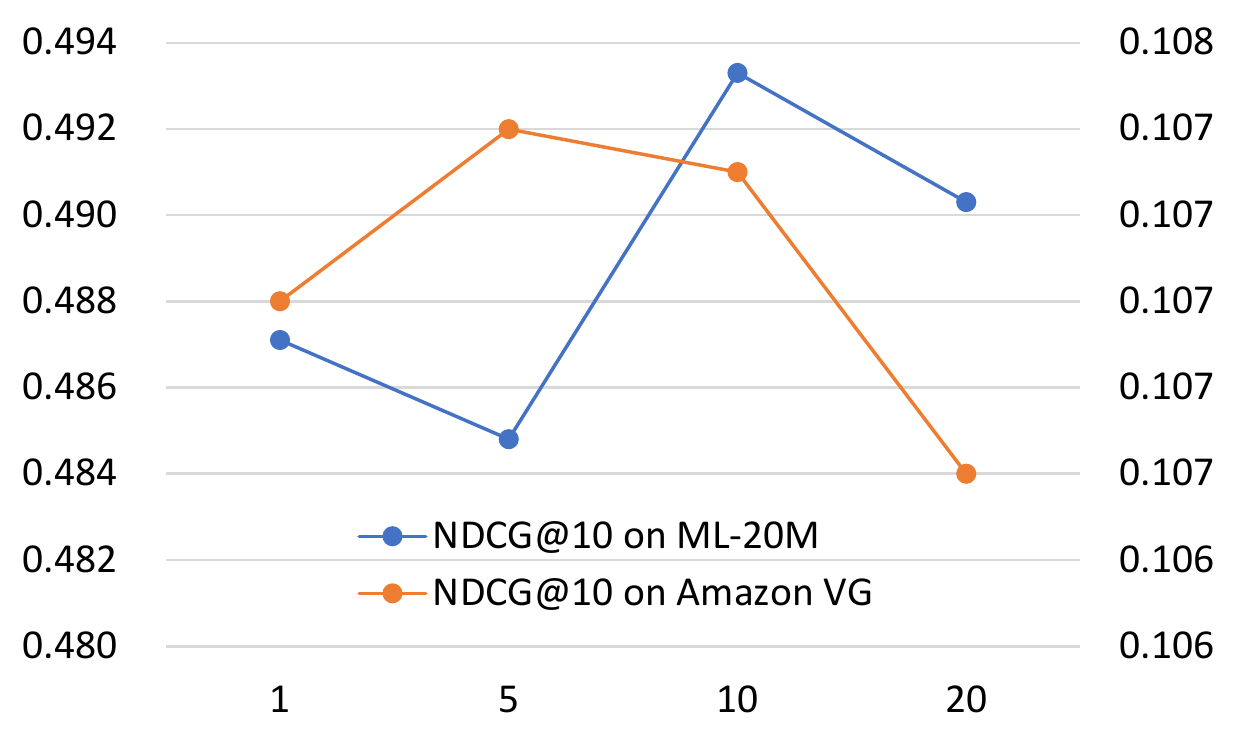}
		\caption{The number of positive samples}
		\label{fig:posNumber}
	\end{subfigure}
	\begin{subfigure}{0.22\textwidth}
		\includegraphics[width=\textwidth]{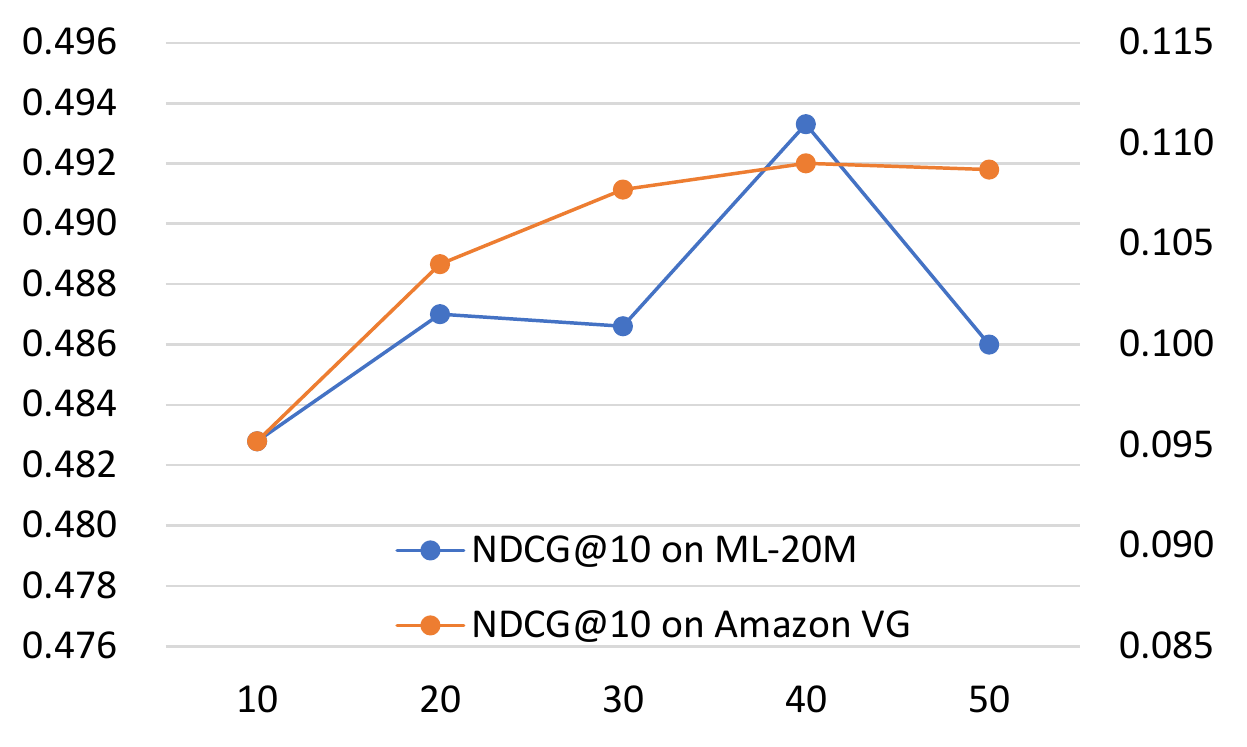}
		\caption{The number of negative samples}
		\label{fig:negNumber}
	\end{subfigure}
	\caption{Tuning of hyper-parameters}
	\label{fig:avatar}
\end{figure}

\subsection{Ablation Study (RQ2)}

Now we investigate the effectiveness of the contrastive CF and the joint training of the two CFs. For this purpose, we compare CCFCRec with its two variants as follows:
\begin{itemize}

\item \textbf{CCFCRec-Co} is the variant without contrastive CF, where $\mathcal{L}_{\text{z}}$ and $\mathcal{L}_{\text{c}}$ are removed from the overall loss (Equation (\ref{eq:loss})).

\item \textbf{CCFCRec-pretrain} is the variant where UCE $\mathbf{s}_u$ and COCE $\mathbf{z}_v$ are pretrained with matrix factorization, and $\mathcal{L}_{\text{z}}$ is removed from the overall loss (Equation (\ref{eq:loss})). During the training, $\mathbf{z}_v$ is fixed.

\end{itemize}

The results are shown in Figure \ref{fig:ablation}. We can see that CCFCRec consistently outperforms the two variants on both datasets. In particular, CCFCRec-Co is significantly inferior to CCFCRec, especially on Amazon-VG, from which we can draw two conclusions: (1)  the collaborative embeddings can be enhanced by contrastive CF indeed; (2) the contrastive CF can achieve greater improvement on sparser dataset, due to its extended contrastive scope from which user preference to item contents can be captured more accurately.

Similarly, CCFCRec-pretrain is also inferior to CCFCRec, which shows the advantage of the joint training of the two CF modules. In fact, the joint training brings two advantages: (1) the two CF modules are trained with the same supervision, which leads to the positive transfer of the co-occurrence collaborative signals; (2) the collaborative embeddings will be adjusted according to the ranking supervision signals, which makes them better adaptable to the ranking task. In contrast, the pretrained embeddings will degrade the performance due to the problem of error propagation.

\begin{figure*}[t!]
	\centering
	\begin{subfigure}{0.23\textwidth}
		\includegraphics[width=\textwidth]{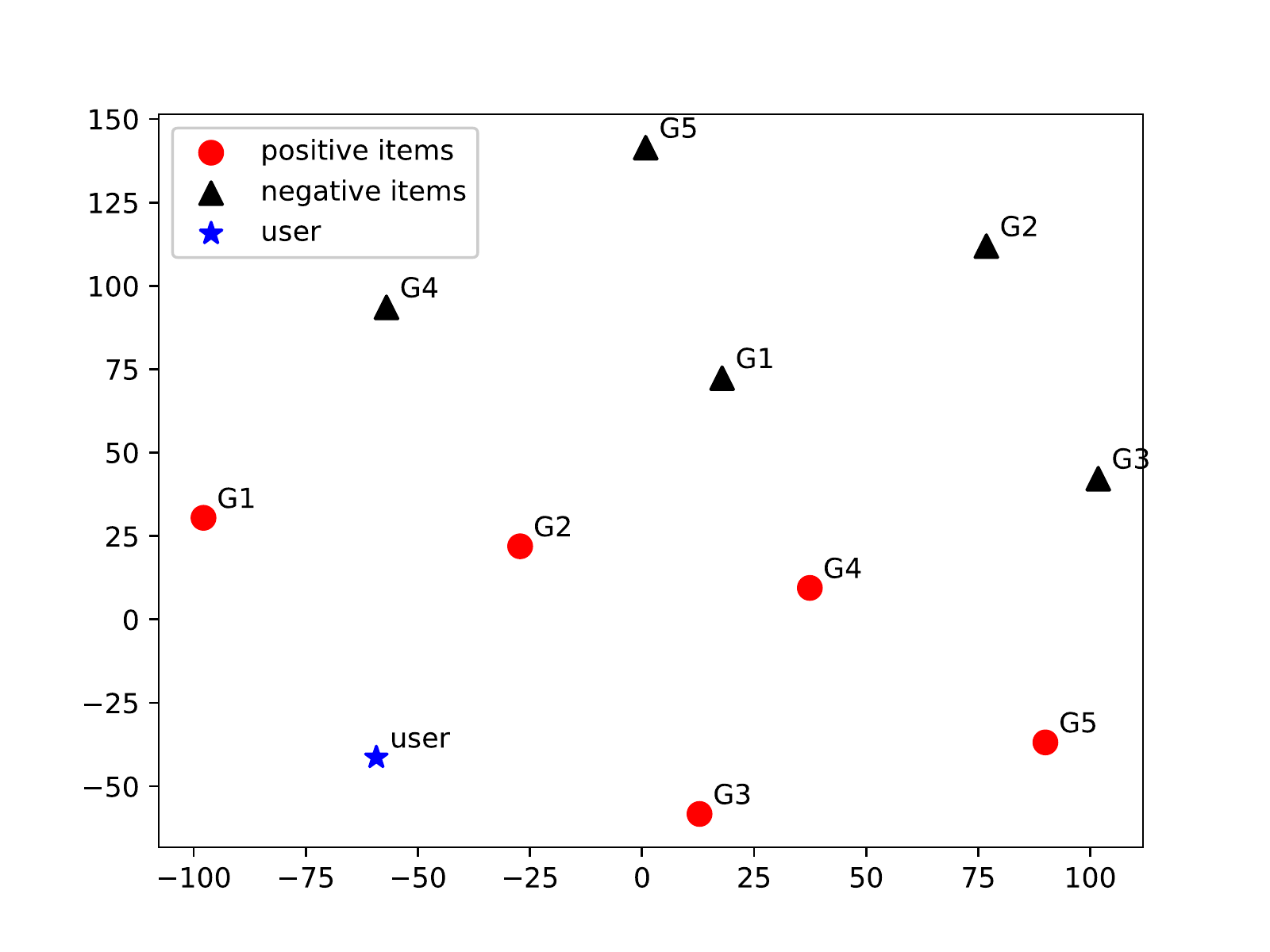}
		\label{fig:user10430fullmodel}
	\end{subfigure}
	\begin{subfigure}{0.23\textwidth}
		\includegraphics[width=\textwidth]{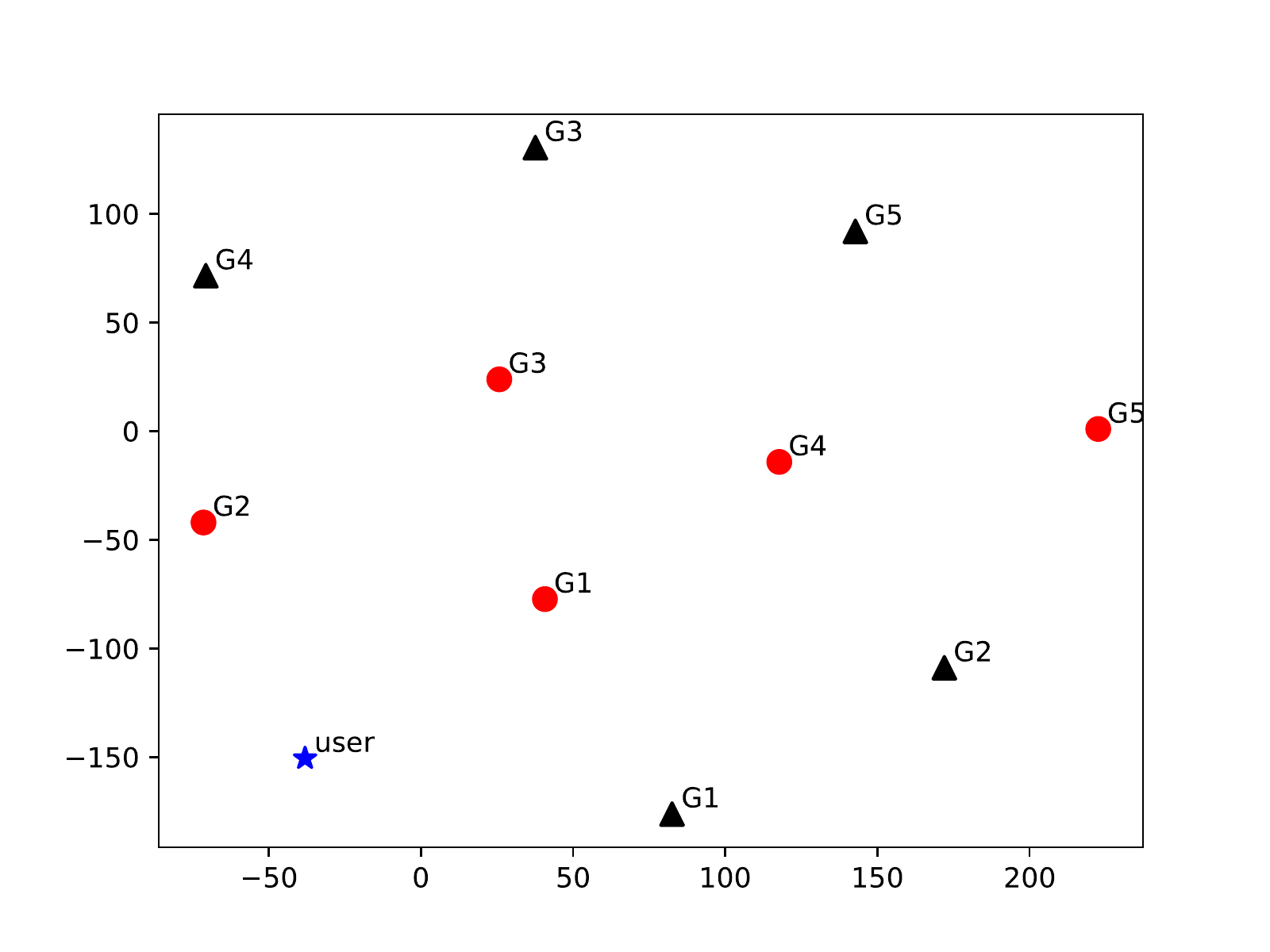}
		\label{fig:user350fullmodel}
	\end{subfigure}
	\begin{subfigure}{0.23\textwidth}
		\includegraphics[width=\textwidth]{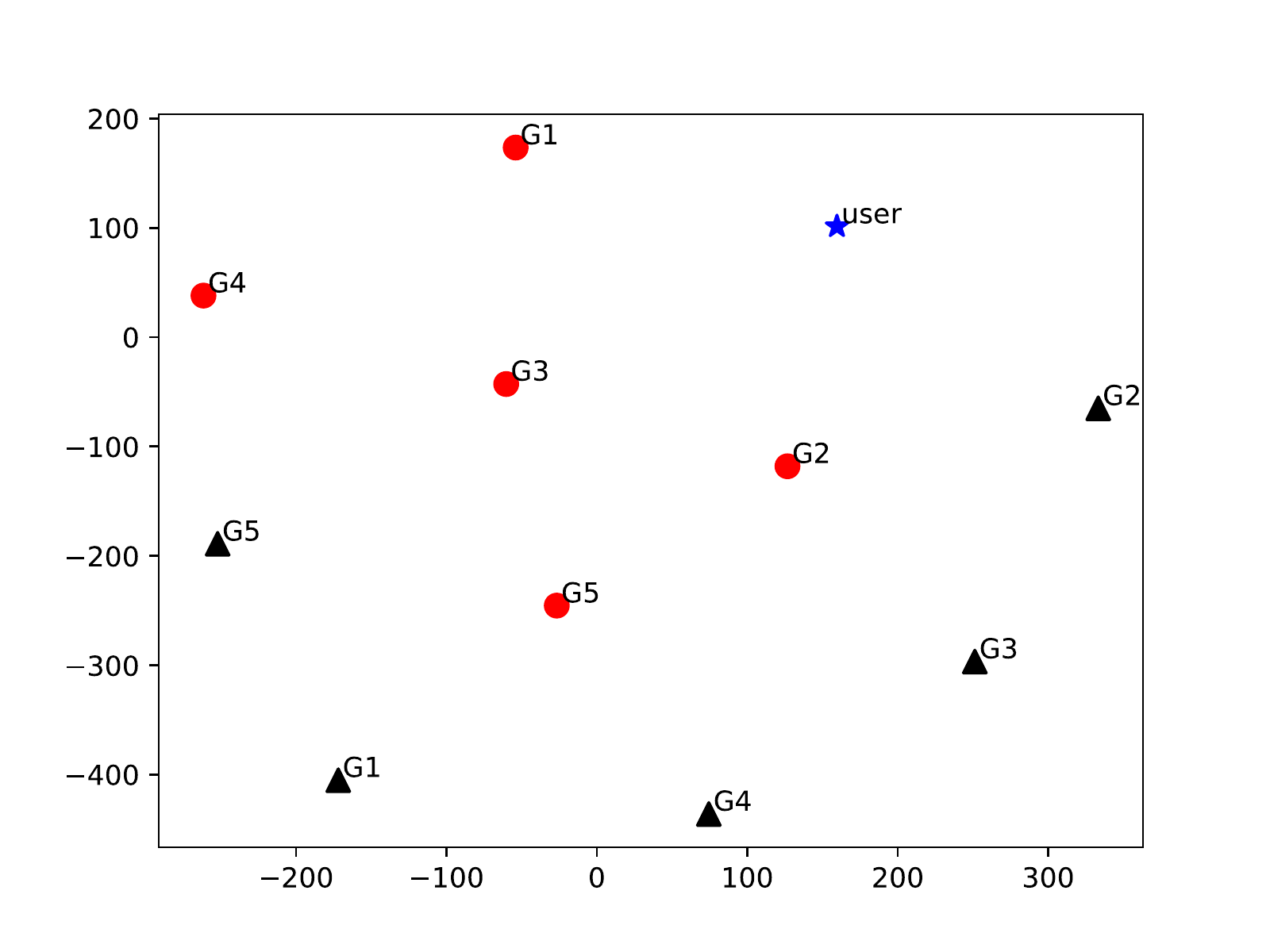}	
		\label{fig:userA1UYRC80M9EZ3Z-CCFC}
	\end{subfigure}
	\begin{subfigure}{0.23\textwidth}
		\includegraphics[width=\textwidth]{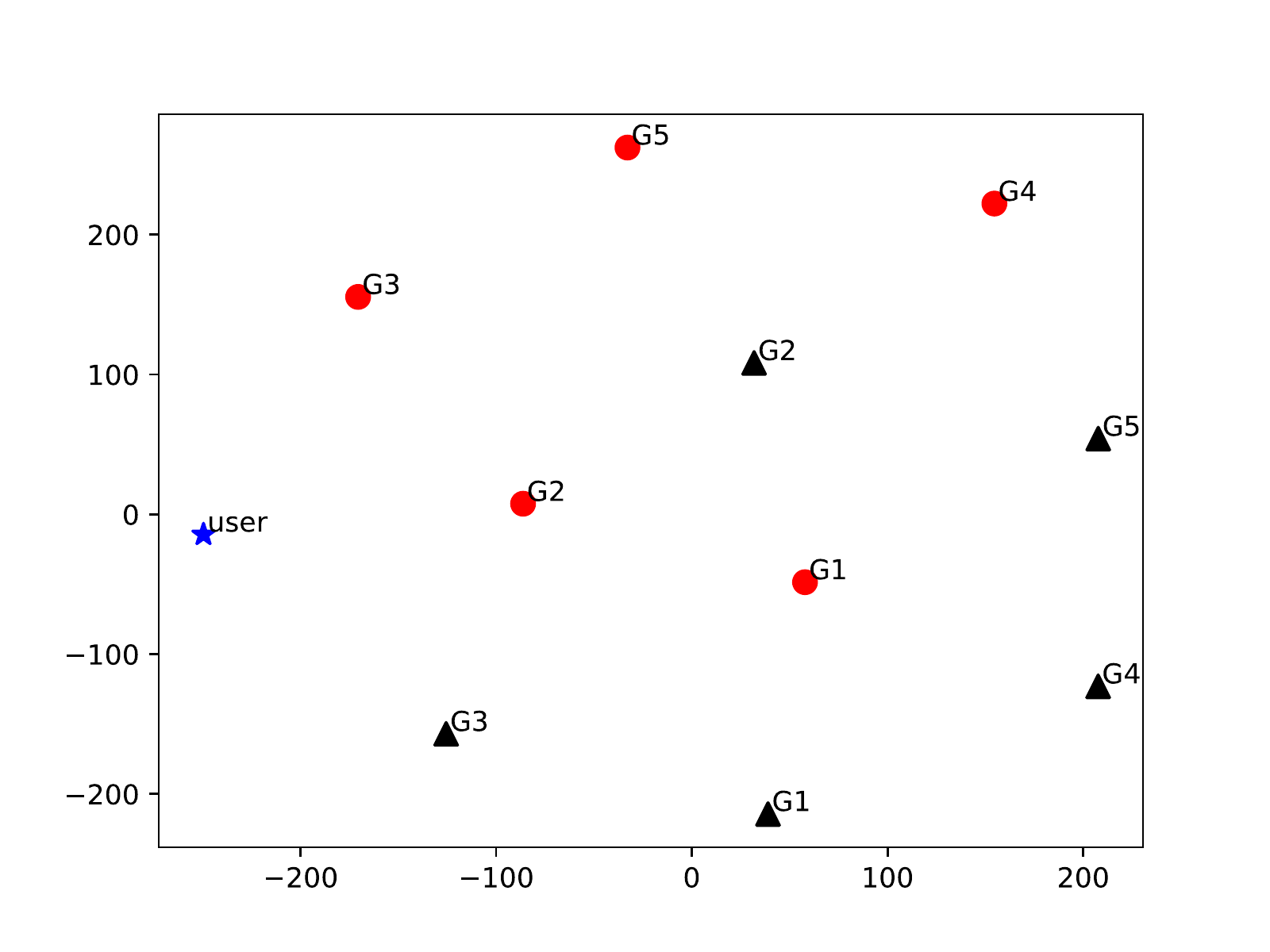}	
		\label{fig:user-A16MO8WTDI3GGB-fullmodel}
	\end{subfigure}
		\begin{subfigure}{0.23\textwidth}
		\includegraphics[width=\textwidth]{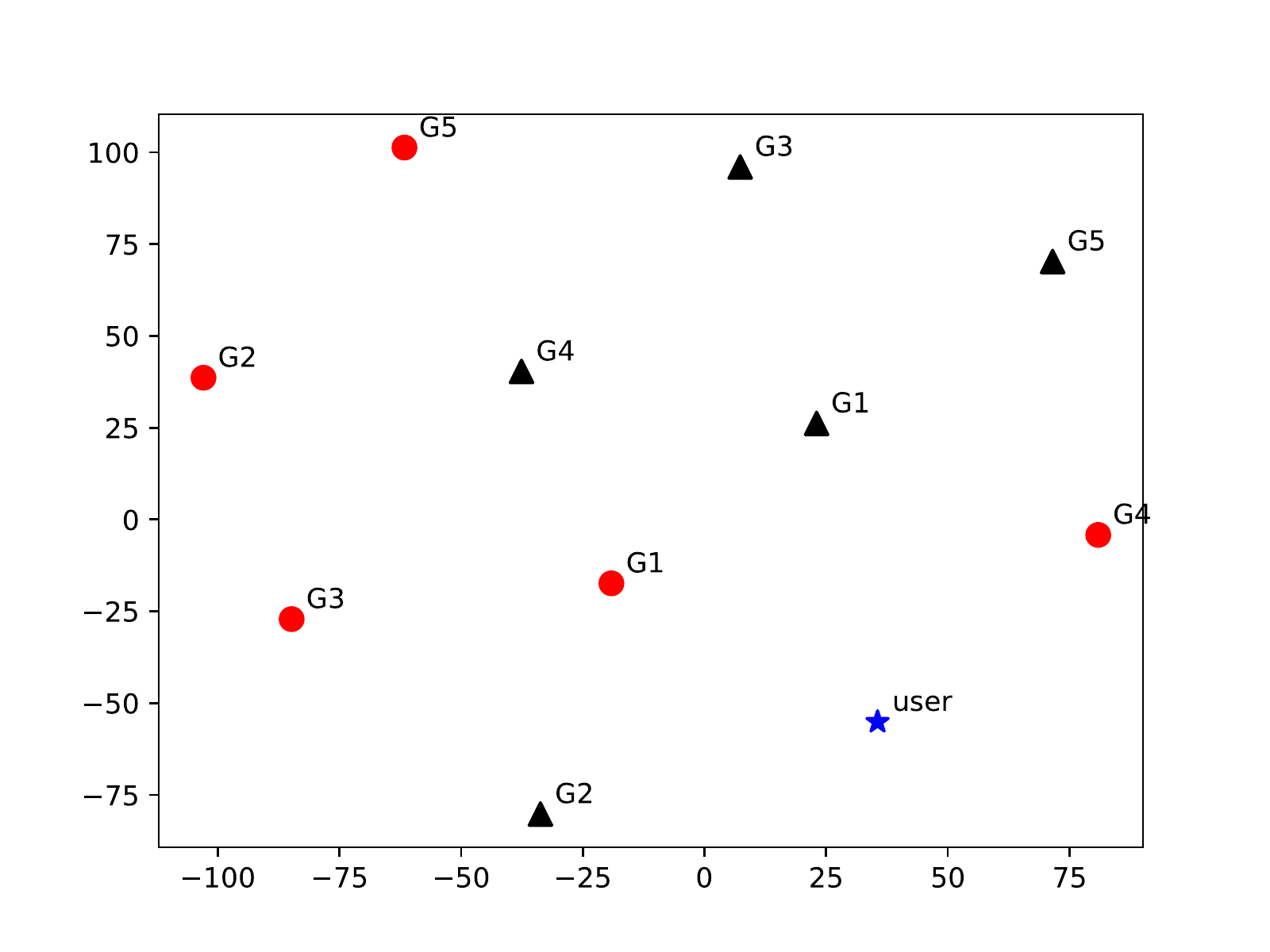}
		\caption{User 10430 in ML-20M}
		\label{fig:user10430variant}
	\end{subfigure}
	\begin{subfigure}{0.23\textwidth}
		\includegraphics[width=\textwidth]{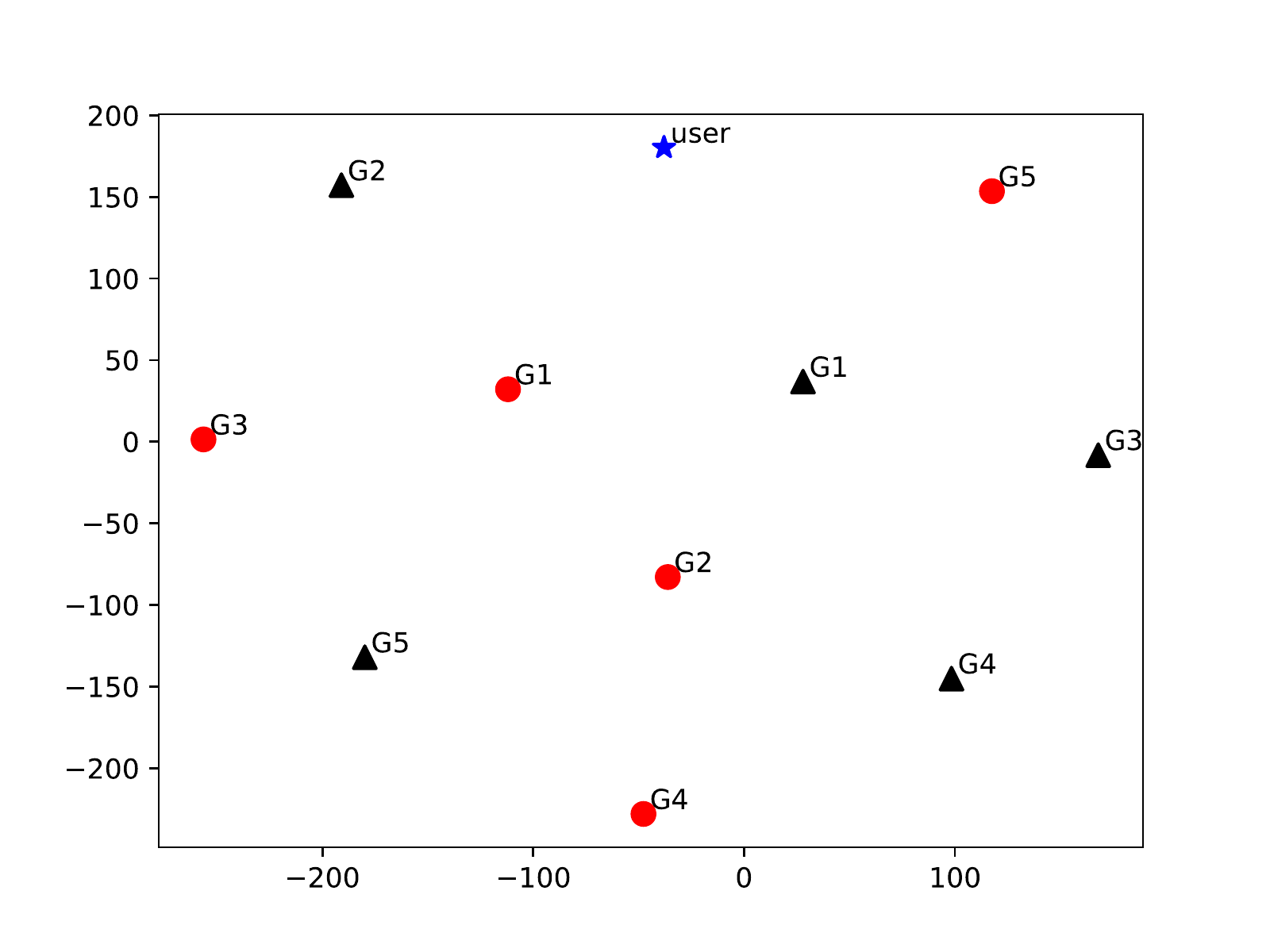}	
		\caption{User 350 in ML-20M}
		\label{fig:user350variant}
	\end{subfigure}	
	\begin{subfigure}{0.23\textwidth}
		\includegraphics[width=\textwidth]{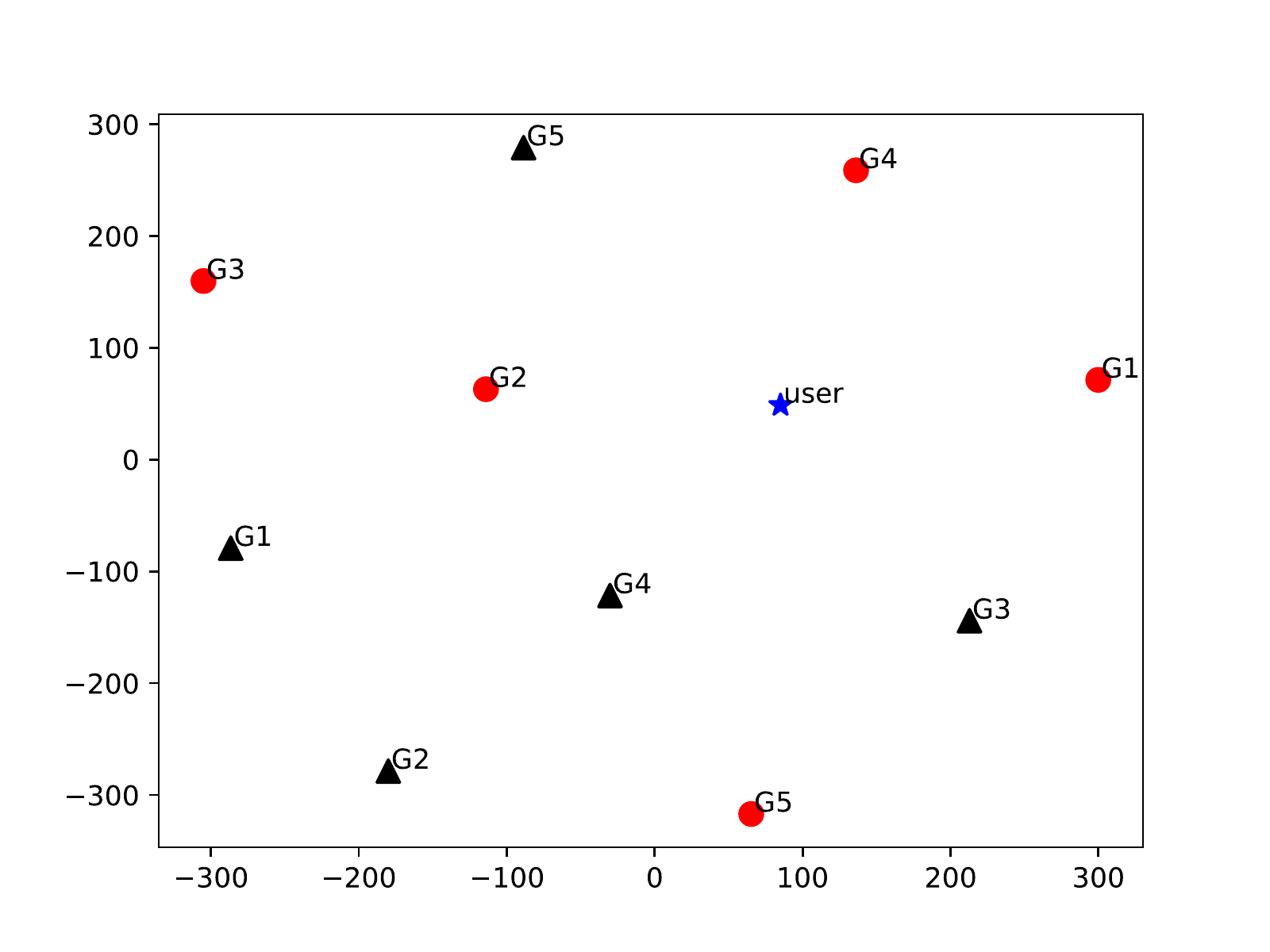}	
		\caption{User A1 in Amazon-VG}
		\label{fig:userA1UYRC80M9EZ3Z-variant}
	\end{subfigure}
		\begin{subfigure}{0.23\textwidth}
		\includegraphics[width=\textwidth]{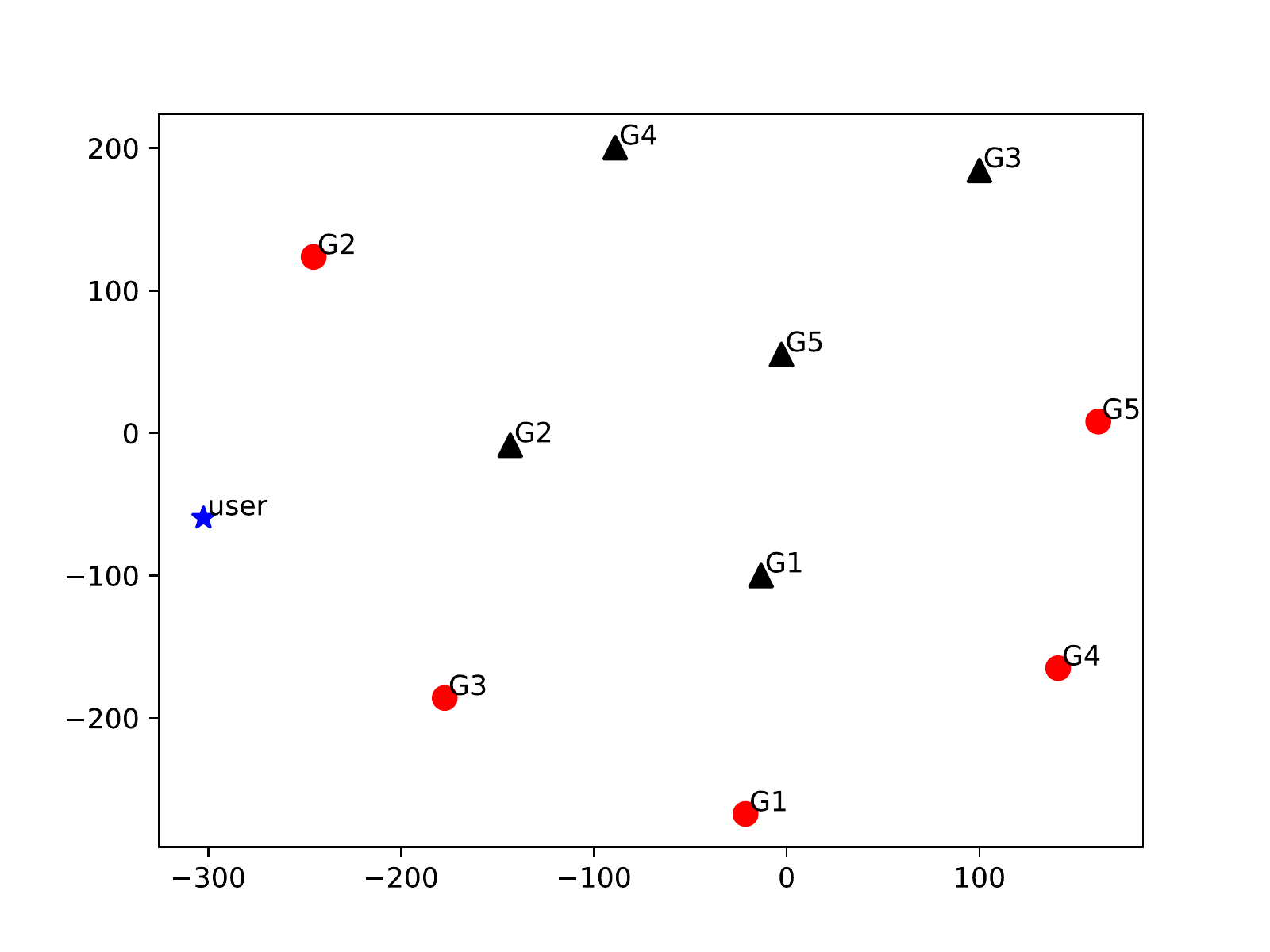}	
		\caption{User A16 in Amazon-VG}
		\label{fig:user-A16MO8WTDI3GGB-variant}
	\end{subfigure}
		
	\caption{Visualization of collaborative embeddings. The first row is the results of CCFCRec, while the second row is the results of CCFCRec-Co. One column corresponds to the results of a user.}
	\label{fig:visual}
\end{figure*}
\begin{table*}[t!]
	 \resizebox{\textwidth}{!}{
	\begin{tabular*}{\linewidth}{@{}c|ll|ll|ll|c|ll|ll|ll@{}}
    \toprule
 \multirowcell{2}{User 10430} & \multicolumn{2}{c|}{$d^+$} & \multicolumn{2}{c|}{$d^-$} & \multicolumn{2}{c|}{$d^- - d^+$} &  
 \multirowcell{2}{User A1} & \multicolumn{2}{c|}{$d^+$} & \multicolumn{2}{c|}{$d^-$} & \multicolumn{2}{c}{$d^- - d^+$}  \\
     & w-Co   & w/o-Co & w-Co   & w/o-Co & w-Co   & w/o-Co &     & w-Co   & w/o-Co & w-Co   & w/o-Co & w-Co   & w/o-Co\\
    \midrule
  G1 & 81.53  & 66.43  & 137.41 & 82.28  & 55.88  & 15.85   &  G1 & 225.42 & 215.99 & 605.28 & 393.04 & 379.86 & 177.05\\
  G2 & 70.97  & 167.23 & 204.98 & 73.66  & 134.01 & -93.57  &  G2 & 221.80 & 199.61 & 240.48 & 421.18 & 18.68  & 221.57\\
  G3 & 74.05  & 123.58 & 181.43 & 153.77 & 107.38 & 30.19   &  G3 & 262.91 & 405.56 & 408.33 & 231.3  & 145.42 & -174.26\\
  G4 & 109.24 & 68.18  & 134.97 & 120.28 & 25.73  & 52.10   &  G4 & 425.83 & 216.19 & 544.03 & 205.57 & 118.20 & -10.62\\
  G5 & 149.3  & 184.17 & 192.45 & 130.46 & 43.15  & -53.71  &  G5 & 393.63 & 366.17 & 503.68 & 288.54 & 110.05 & -77.63 \\
	\bottomrule
    \multirowcell{2}{User 350} & \multicolumn{2}{c|}{$d^+$} &  \multicolumn{2}{c|}{$d^-$} & \multicolumn{2}{c|}{$d^- - d^+$} &  
    \multirowcell{2}{User A16} & \multicolumn{2}{c|}{$d^+$} &  \multicolumn{2}{c|}{$d^-$} & \multicolumn{2}{c}{$d^- - d^+$} \\
       & w-Co   & w/o-Co & w-Co   & w/o-Co & w-Co   & w/o-Co &     & w-Co   & w/o-Co & w-Co   & w/o-Co & w-Co   & w/o-Co\\
    \midrule
    G1 & 107.61 & 165.54 & 123.27 & 157.78 & 15.66  & -7.76   & G1 & 309.23 & 349.56 & 350.92 & 291.95 & 41.69  & -57.61\\
    G2 & 113.55 & 263.17 & 214.17 & 154.74  &100.62 & -108.43 & G2 & 164.77 & 191.98 & 306.95 & 167.13 & 142.18 & -24.85\\
    G3 & 185.73 & 282.29 & 291.10  & 279.16 & 105.37 & -3.13   & G3 & 187.46 & 177.79 & 188.89 & 470.58 & 1.43   & 292.79 \\
    G4 & 207.10 & 408.44 & 224.53 & 352.76 & 17.43  & -55.68  & G4 & 468.36 & 455.57 & 469.81 & 336.45 & 1.45   & -119.12 \\
    G5 & 301.51 & 157.82 & 302.54 & 342.99 & 1.03  & 185.17   & G5 & 351.52 & 468.95 & 462.30 & 321.04 & 110.78 & -147.91\\
	\bottomrule  
	\end{tabular*}
}

\caption{The distances from item embedding to the user embedding. w-Co represents CCFCRec, while w/o-Co CCFCRec-Co. G1-G5 represent five pairs of positive-negative sample. $d^+$ represents the positive samples' distances to the user, while $d^-$ the negative samples' distances.}
\label{tbl:distance}
\end{table*}

\subsection{Tuning of Hyper-parameter (RQ3)}
In this section, on the validation sets of both datasets, we investigate the influence of hyper-parameters, the embedding dimensionality $d$, the balance factor of the contrastive loss $\lambda$, the numbers of positive samples and negative samples for a training item, of which the results are shown in Figure \ref{fig:avatar}.

The curves of the embedding dimensionality $d$ are shown in Figure \ref{fig:avatar}(a), from which we observe that NDCG@10 reaches the peak at $d = 128 $ on ML-20M and $d = 256$ on Amazon-VG, respectively.

From Figure \ref{fig:avatar}(b) we can see that on both datasets, the performance of CCFCRec increases first with the increasing of $\lambda$, which shows that the model performance will improve as more co-occurrence collaborative signals are captured by the contrastive CF of CCFCRec. However, when $\lambda$ is over $0.5$ on ML-20M and $0.6$ on Amazon-VG, the model performance begins to drop, because excessive co-occurrence collaborative signals (due to the too small weight of $\mathcal{L}_{\text{q}}$ and $\mathcal{L}_{\text{z}}$) and insufficient content-based collaborative signals (due to the too big weight of $\mathcal{L}_{\text{c}}$) are captured.

Similarly, as shown in Figures \ref{fig:avatar}(c) and \ref{fig:avatar}(d), the model performance curves first rise and then fall with the increasing of the numbers of the positive and negative samples. Obviously, more positive and negative samples for a training item benefits the contrastive learning for the transfer of the co-occurrence collaborative signals, which explains the rising part. However, excessive contrastive samples may incur the risk of overfitting.

\subsection{Case Study (RQ4)}
Now we visualize the collaborative embeddings learned by CCFCRec and CCFCRec-Co by case studies. We randomly select two users from ML-20M (User 10430 and User 350) and Amazon-VG (User A1 and User A16), respectively. For each user,  we pair each of the top-5 positive samples to a randomly chosen negative sample with the same genre or category attribute, and plot the UCEs and the CBCEs with t-SNE algorithm \cite{van2008visualizing} in Figure \ref{fig:visual}. We also show the distances of item embeddings to corresponding user embeddings in Table \ref{tbl:distance}. 

We can observe that for each pair of samples, the positive sample embedding learned by CCFCRec is closer to the user embedding than the negative one as shown in the first row of Figure \ref{fig:visual} and the w-Co of $d^- - d^+$ columns in Table \ref{tbl:distance}, while the embeddings learned by CCFCRec-Co are blurred and cannot keep the proper ranking order, as shown in the second row of Figure \ref{fig:visual} and the w/o-Co of $d^- - d^+$ columns in Table \ref{tbl:distance}. For example, the bottom figure in Figure \ref{fig:visual}(a) together with the cell (User 101430, G2, $d^--d^+$, w/o-Co) in Table \ref{tbl:distance} shows that in the G2 pair of User 101430, the negative sample embedding is closer to the user embedding than the positive one by 93.57. In contrast, the positive sample embedding learned by CCFCRec is closer than the negative one by 134.01 (as shown in the cell (User 101430, G5, $d^--d^+$, w-Co) in Table \ref{tbl:distance}). This result demonstrates the superiority of the contrastive CF of CCFCRec in ranking the positive samples ahead of negative ones, due to its ability to leverage the co-occurrence collaborative signals to rectify the blurry collaborative embeddings.

\section{Related works}
\subsection{Cold-start Recommendation}
The existing works for cold-start recommendation often utilize side information to overcome the problem of sparse data. Content-based  works usually capture collaborative signals from item content so that the collaborative filtering can be applicable for cold-start items. For example, NSPR \cite{ebesu2017neural} employs DNN to extract content features and use collaborative filtering to predict the user ratings to cold-start items. Different from directly modeling item content, recent content-based methods learn the correlation between side information and item behavior features, and train a generative model to project a cold item's content onto a warm item embedding space \cite{gantner2010learning, saveski2014item,mo2015image,volkovs2017dropoutnet,Barkan2019,Li2019,Pan2019,sun2020lara,Zhu2021,chen2022generative}. For example, BPR-KNN \cite{gantner2010learning} employs KNN and linear function to constraint item content feature and item behavior feature. LCE \cite{saveski2014item} combines content information and collaboration information into a common low-dimensional space. LARA \cite{sun2020lara} exploits an adversarial neural network with multiple generators to learn a mapping from cold-start items' attributes to user embeddings to generate virtual users for cold-start items.

Some efforts tackle the cold-start problem from the perspective of robust learning, which treat cold-start items as warm-start items missing interaction records, and try to infer the warm embeddings of the cold-start items in a robust manner like randomly corrupting the collaborative embedding of warm-start training items. For example, DropoutNet \cite{volkovs2017dropoutnet} selects a subset of users/items during training, and removes their behavior features to simulate the cold-start scenario.
MTPR \cite{du2020learn} constructs a counterfactual representation by setting the behavior part of an item's latent representation to zero, and learns pairwise matching between normal representations and counterfactual representations with BPR loss.

Some works use meta-learning to address cold-start recommendation. For example, MeLU \cite{lee2019melu} uses two sets, namely, the support set and query set. The support set and query set are used for calculating training loss and test loss for each recommendation task. During the local update phase, the model minimizes the training loss based on the support set, while during the global update phase, the model minimizes the test loss based on the query set. MetaHIN \cite{lu2020meta} extends meta-learning into heterogeneous information networks to address cold-start recommendations. MWUF \cite{Zhu2021} generates warm embeddings for cold-start items based on their features and ID embeddings with two meta networks.

\subsection{Contrastive Learning in Recommendation}

Contrastive Learning (CL) is a self-supervised technique, which has achieved an impressive success in solving the problem of data sparsity \cite{chen2020simple,henaff2020data,yan2021consert}. Recently, some studies explore contrastive learning for recommender systems \cite{yu2022self,ma2020disentangled,qiu2021memory,wu2021self,xia2021self,yu2021socially,yu2021self,zhou2020s3,Yu2022,xie2022contrastive,chen2022intent}. For example, Zhou \textit{et al.} \cite{zhou2020s3} apply a random masking of attributes and items to build different views and learn the correlations among attributes, items, subsequences, and sequences by maximizing the mutual information between the different views.
SGL \cite{wu2021self} models user-item interactions as a graph and builds multiple views by conducting node dropout, edge dropout, and random walk on the user-item graph. Similarly, Similarly, Yu \textit{et al.} \cite{Yu2022} propose a simple CL method to perturb the graph embedding space with uniform noise to build contrastive views for the training of recommendation model. At the same time, some researchers model user-item interactions with hypergraph, and conduct CL over augmented hypergraph views for the user/item representation learning \cite{yu2021self,xia2021self,zhang2021double}. Wei \textit{et al.} \cite{wei2021contrastive} propose a cold-start recommendation model CLCRec, which fulfills the learning of feature representations for cold-start items by maximizing the mutual information between item content and collaborative signals.

Different from the existing works which often limit the contrastive learning to the views of the same item, our CCFCRec extends the contrastive scope to the views of the second-order neighbors of a training item, which helps capture more collaborative signals and alleviate the sparsity problem as well.

\section{Conclusion}

In this paper, we propose a novel model called Contrastive Collaborative Filtering for Cold-start item recommendation (CCFCRec). CCFCRec can take advantage of the co-occurrence collaborative signals in warm training data to alleviate the issue of blurry collaborative embeddings. In particular, we devise a contrastive collaborative filtering (CF) framework, consisting of a content CF module and a co-occurrence CF module to generate the content-based collaborative embedding and the co-occurrence collaborative embedding for a training item, respectively. During the joint training of the two CF modules, the contrastive learning between the two collaborative embeddings enables the content CF module to memorize the co-occurrence collaborative signals, so that during the applying phase, the blurry collaborative embeddings can be rectified. The extensive experiments conducted on real datasets together with the sound theoretical analysis demonstrate the superiority of CCFCRec.

\begin{acks}
This work is supported by National Natural Science Foundation of China under grant 61972270.
\end{acks}
\bibliographystyle{ACM-Reference-Format}
\bibliography{CCFCRec}

\newpage
\appendix
\section{Appendix}
\begin{table}[h]
	\begin{center}
		\caption{Hyper-parameter setting of baseline methods. } 
		\label{tab:table2}
		\begin{tabular}{c|c|c|c} 
			\hline
			\multirow{2}{*}{\textbf{Baseline}} & \multicolumn{2}{c|}{\textbf{\makecell{Embedding dimensionality \\ selected from  \\ \{8, 16, 32, 64, 128, 256\} } }}  &  
\multirow{2}{*}{\textbf{Other hyper-parameters}}
\\
\cline{2-3}
                            & ML-20M & Amazon VG &  \\
                \hline
			NFM         & 64  & 64      & \makecell{Number of hidden layers = 2 selected from \{1, 2, 3\} }                      \\
			\hline
			LARA        & 128 & 64      & \makecell{Number of hidden layers \\ in the Generator and Discriminator = 3 \\selected from \{1, 2, 3, 4\} }                      \\
			\hline
			MTPR        & 64  & 128     & \multirow{2}{*}{/}                      \\
                \cline{1-3}
			CVAR        & 64  & 64      &                       \\
			\hline
			MvDGAE      & 128 & 64     & \makecell{Dropout Rate=0 \\ selected from \{0, 0.2, 0.4, 0.6, 0.8, 1.0\} }      \\
			\hline
			\multirow{2}{*}[-2.0ex]{CLCRec}      & \multirow{2}{*}[-2.0ex]{64}  & \multirow{2}{*}[-2.0ex]{256}    & \makecell{Number of negative samples k=128 \\selected from \{16, 32, 64, 128, 256\}  }    \\
			\cline{4-4}
                         &     &        & \makecell{The weight of $\mathcal{L}_{RE}$  $\lambda$=0.5 \\ selected from \{0, 0.1, 0.2, ..., 1.0\}} \\
			\hline
		\end{tabular}
	\end{center}
\end{table}

\end{document}